\documentclass[sigconf]{acmart}

\copyrightyear{2022}
\acmYear{2022}
\setcopyright{acmcopyright}\acmConference[ICSE '22]{44th International Conference on Software Engineering}{May 21--29, 2022}{Pittsburgh, PA, USA}
\acmBooktitle{44th International Conference on Software Engineering (ICSE '22), May 21--29, 2022, Pittsburgh, PA, USA}
\acmPrice{15.00}
\acmDOI{10.1145/3510003.3510041}
\acmISBN{978-1-4503-9221-1/22/05}

\AtBeginDocument{%
  \providecommand\BibTeX{{%
    \normalfont B\kern-0.5em{\scshape i\kern-0.25em b}\kern-0.8em\TeX}}}

\usepackage{amsmath,amsfonts}
\usepackage{algorithmic}
\usepackage[ruled,linesnumbered,vlined]{algorithm2e}
\usepackage{graphicx}
\usepackage{textcomp}
\usepackage{xcolor}
\usepackage{url}
\usepackage{xspace}
\usepackage{balance}
\usepackage{booktabs, multirow} 
\usepackage{listings}
\begin{document}
\newcommand{\precaptionspace}{\vspace{-3ex}}
\newcommand{\pretabcaptionspace}{\vspace*{-1.5ex}}
\newcommand{\posttabcaptionspace}{\vspace*{-1.5ex}}
\newcommand{\Comment}[1]{}
\newcommand{\Space}[1]{}
\newcommand{\parabf}[1]{\noindent\textbf{#1}}
\newcommand{\smallspace}{\hspace{0.5em}}
\newcommand{\ccc}[1]{\hfill \begin{small}\# #1\end{small}}
\newcommand{\Num}[1]{\textbf{#1}}
\newcommand{\CodeIn}[1]{\begin{small}\texttt{#1}\end{small}}
\newcommand{\Fix}[1]{{\color{red}\bfseries [#1]}}
\newcommand{\lingming}[1]{{\color{red}\bfseries [#1]}}
\newcommand{\david}[1]{{\color{brown}\bfseries [#1]}}
\newcommand{\yinlin}[1]{{\color{purple}\bfseries [#1]}}
\newcommand{\chenyuan}[1]{{\color{blue}\bfseries [#1]}}

\newcommand{\tech}{FreeFuzz\xspace} 
\newcommand{\pt}{PyTorch\xspace}
\newcommand{\tf}{TensorFlow\xspace}
\newcommand{\keras}{Keras\xspace}
\newcommand{\fuzztype}{\CodeIn{FuzzType}\xspace}
\newcommand{\fuzztypenocode}{FuzzType\xspace}
\newcommand{\cradle}{CRADLE\xspace}
\newcommand{\lemon}{LEMON\xspace}
\newcommand{\predoo}{Predoo\xspace}
\newcommand{\convtwod}{2D-Convolution\xspace}
\newcommand{\convthreed}{3D-Convolution\xspace}
\newcommand{\apivs}{API value space\xspace}
\newcommand{\vs}{value space\xspace}
\newcommand{\argdb}{argument value space\xspace}
\newcommand{\tensordim}{Tensor Dim Mutation\xspace}
\newcommand{\tensordtype}{Tensor Dtype Mutation\xspace}
\newcommand{\typetransform}{Primitive Mutation\xspace}
\newcommand{\tuplemutation}{Tuple Mutation\xspace}
\newcommand{\listmutation}{List Mutation\xspace}
\newcommand{\randomtensorshape}{Random Tensor Shape\xspace}
\newcommand{\randomtensorvalue}{Random Tensor Value\xspace}
\newcommand{\randomprimitive}{Random Primitive\xspace}
\newcommand{\randomtuple}{Random Tuple\xspace}
\newcommand{\randomlist}{Random List\xspace}
\newcommand{\dbtensorshape}{Database Tensor Shape\xspace}
\newcommand{\dbtensorvalue}{Database Tensor Value\xspace}
\newcommand{\dbprimitive}{Database Primitive\xspace}
\newcommand{\dbtuple}{Database Tuple\xspace}
\newcommand{\dblist}{Database List\xspace}

\title{Free Lunch for Testing:\\ Fuzzing Deep-Learning Libraries from Open Source\Comment{ via API Instrumentation}}

\newcommand{\NumAllBugs}{49\xspace} 
\newcommand{\NumPreviouslyUnknownConfirm}{38\xspace} 
\newcommand{\NumAllBugsPt}{28\xspace}
\newcommand{\NumAllBugsTf}{21\xspace}
\newcommand{\NumConfirmedBugsPt}{23\xspace} 
\newcommand{\NumConfirmedBugsTf}{15\xspace} 
\newcommand{\numFixedBugs}{21\xspace}
\newcommand{\numFixedPt}{7\xspace}
\newcommand{\numFixedTf}{14\xspace}
\newcommand{\numTracedAPIs}{1158\xspace}
\newcommand{\numTracedAPIsPt}{470\xspace}
\newcommand{\numTracedAPIsTf}{688\xspace}
\newcommand{\numHookAPIs}{2530\xspace}
\newcommand{\numHookAPIsPt}{630\xspace}
\newcommand{\numHookAPIsTfone}{2888\xspace}
\newcommand{\numHookAPIsTftwo}{1900\xspace}
\newcommand{\lemonModelNum}{12\xspace}
\newcommand{\cradleModelNum}{30\xspace}
\newcommand{\cradleDatasetNum}{11\xspace}
\newcommand{\predooOperatorNum}{7\xspace}
\newcommand{\numModelsPtAll}{102\xspace}
\newcommand{\numModelsTfAll}{100\xspace}
\newcommand{\numModelsAll}{202\xspace}

\newcommand{\numdocpt}{497\xspace}
\newcommand{\numtestpt}{1493\xspace}
\newcommand{\numdoctf}{512\xspace}
\newcommand{\numtesttf}{1420\xspace}

\newcommand{\numdocapipt}{427\xspace}
\newcommand{\numdocapivspt}{1259\xspace}
\newcommand{\numtestapipt}{176\xspace}
\newcommand{\numtestapivspt}{3383\xspace}
\newcommand{\numtestapitf}{216\xspace}
\newcommand{\nummodelapipt}{145\xspace}
\newcommand{\nummodelapitf}{269\xspace}

\newcommand{\overheadlowerlemonmodel}{75X\xspace}

\newcommand{\overheadlowerfull}{3.5X\xspace}

\newcommand{\APImorecovered}{9X\xspace}

\newcommand{\morecoveragelemon}{20\%\xspace}

\newcommand{\numcradleAPI}{59\xspace}
\newcommand{\numlemonAPI}{35\xspace}

\newcommand{\linecovFreeComp}{35473\xspace}
\newcommand{\linecovlemon}{29766\xspace}

\newcommand{\lbracket}{\langle}
\newcommand{\rbracket}{\rangle}

\captionsetup[figure]{font=bf,skip=0.2em} 
\captionsetup[table]{font=bf,skip=0.2em} 
\newcommand{\distance}{0.5em}
\setlength{\textfloatsep}{\distance} 
\setlength{\floatsep}{\distance} 
\setlength{\intextsep}{\distance} 
\setlength{\dbltextfloatsep}{\distance} 
\setlength{\dblfloatsep}{\distance} 


\author{Anjiang Wei}\authornote{The work was done during a remote summer internship at University of Illinois.}

\affiliation{%
  \institution{Stanford University}
 \country{}
}
\email{anjiang@stanford.edu}

\author{Yinlin Deng}
\affiliation{%
  \institution{University of Illinois at Urbana-Champaign}
  \country{}
}
\email{yinlind2@illinois.edu}

\author{Chenyuan Yang}\authornotemark[1]
\affiliation{%
  \institution{Nanjing University}
  \country{}
}
\email{cy1yang@outlook.com}

\author{Lingming Zhang}
\affiliation{%
 \institution{University of Illinois at Urbana-Champaign}
 \country{}
}
\email{lingming@illinois.edu}





\renewcommand{\shortauthors}{Anjiang Wei, Yinlin Deng, Chenyuan Yang, and Lingming Zhang}

\begin{abstract}
Deep learning (DL) systems can make our life much easier, and thus are gaining more and more attention from both academia and industry. Meanwhile, bugs in DL systems can be disastrous, and can even threaten human lives in safety-critical applications. 
To date, a huge body of research efforts have been dedicated to testing DL models. However, interestingly, there is still limited work for testing the underlying DL libraries, which are the foundation for building, optimizing, and running DL models. 
One potential reason is that test generation for the underlying DL libraries can be rather challenging since their public APIs are mainly exposed in Python, making it even hard to automatically determine the API input parameter types due to dynamic typing. 
In this paper, we propose \tech{}, the first approach to fuzzing DL libraries via mining from open source. More specifically, \tech{} obtains code/models from three different sources: 1) code snippets from the library documentation, 2) library developer tests, and 3) DL models in the wild. Then, \tech{} automatically runs all the collected code/models with instrumentation to trace the dynamic information for each covered API, including the types and values of each parameter during invocation, and shapes of input/output tensors. Lastly, \tech{} will leverage the traced dynamic information to perform fuzz testing for each covered API. 
The extensive study of \tech{} on \pt and \tf, two of the most popular DL libraries, shows that \tech{} is able to automatically trace valid dynamic information for fuzzing \numTracedAPIs\Comment{ of \numHookAPIs} popular APIs, \APImorecovered more than state-of-the-art \lemon with \overheadlowerfull lower overhead than \lemon. To date,\Comment{ leveraging the traced information,} \tech{} has detected \NumAllBugs bugs for \pt and \tf (with \NumPreviouslyUnknownConfirm already confirmed by developers as previously unknown)\Comment{\lingming{not much diff for 42 or 43 (both are very high, excellent job!), let's simply say ``(with 42 already confirmed as previously unknown bugs)''; same change for intro }}\Comment{the numbers don't match the result table, please check all numbers globally}\Comment{, substantially outperforming state-of-the-art DL library testing techniques}\Comment{ \david{by covering \APImorecovered more the number of APIs, \morecoveragelemon more line coverage, and \overheadlowerfull lower overhead}}.

\end{abstract}

\maketitle

\section{Introduction}
Deep Learning (DL) has been playing a significant role in various application domains, including image classification~\cite{He_2016_CVPR,simonyan2014very}, natural language processing~\cite{graves2013speech,gers2000learning}, game playing~\cite{silver2016mastering}, and software engineering~\cite{li2019deepfl,chen2020enhanced,zeng2021deep,yang2020survey}\Comment{\lingming{is ~\cite{ni2021soar} really a deep learning based tech?}\david{It uses NLP models for error message understanding. I am fine to remove it}}.
Through such applications, DL has substantially improved our daily life~\cite{wu2016google,sun2015deepid3,shen2017deep,grigorescu2020survey,bojarski2016end}.
\Comment{In such application domains, DL has been applied to various important tasks (such as machine translation~\cite{wu2016google}, face recognition~\cite{sun2015deepid3}, medical diagnosis~\cite{shen2017deep}, and autonomous driving~\cite{grigorescu2020survey,bojarski2016end}), and has substantially improved our daily life.} 
The great success achieved by DL is attributed to the proposal of more and more advanced DL models, the availability of large-scale datasets, and inevitably, the continuous development of DL libraries. Meanwhile, deploying a DL model without thorough testing can be disastrous in safety-critical applications. For example, a critical bug in the DL system in Uber’s self-driving cars has unfortunately taken the life of a pedestrian~\cite{uberkill}.

Due to the popularity of DL models and the critical importance of their reliability, a growing body of research efforts have been dedicated to testing DL models, with focus on adversarial attacks~\cite{goodfellow2014explaining,moosavi2016deepfool,carlini2019evaluating,madry2017towards,akhtar2018threat,papernot2016limitations} for model robustness, the discussion on various metrics for DL model testing~\cite{harel2020neuron,pei2017deepxplore,yan2020correlations,kim2019guiding,ma2018deepgauge}, and testing DL models for specific applications~\cite{tian2018deeptest,zhang2018deeproad,zhou2020deepbillboard}.
Meanwhile, both running and testing DL models inevitably involve the underlying DL libraries, which serve as central pieces of infrastructures for building, training, optimizing and deploying DL models.\Comment{Nowadays, the most two popular DL libraries are \pt~\cite{paszke2019pytorch} and \tf~\cite{abadi2016tensorflow}. Given the importance of DL libraries, there should be a huge interest in implementing them correctly.}
For example, the popular \pt and \tf DL libraries, with 50K and 159K stars on GitHub, are by far two of the most popular DL libraries for developing and deploying DL models.
Surprisingly, despite the importance of DL library reliability, there is only limited work for testing DL libraries to date. For example, \cradle\cite{cradle} leverages existing DL models for testing \keras\cite{keras} and its backends, and resolves the test oracle problem via differential testing\Comment{ and inconsistency detection techniques }. Later, \lemon\cite{lemon} further augments \cradle via leveraging various model mutation rules to generate more diverse DL models to invoke more library code to expose more possible DL library bugs\Comment{, guided by their heuristic strategies with the goal of amplifying inconsistencies between different libraries}.
\Comment{More recently, the \predoo work~\cite{zhang2021predoo} proposes an operator-level precision testing approach for \predooOperatorNum DL operators from \tf to maximize their precision errors.\lingming{need to better explain why it does not quite work; otherwise, it is strange we dont compare with it}}

Despite their promising results, existing work on testing DL libraries suffers from the following limitations. Firstly, only limited sources for test input generation are considered. For example, \Comment{\predoo~\cite{zhang2021predoo} manually constructs the test input\lingming{i don't quite understand here, can you elaborate}, i.e. selecting only \predooOperatorNum operators from \tf, which is not scalable. }\cradle~\cite{cradle} uses \cradleModelNum pre-trained DL models and \lemon~\cite{lemon} uses only \lemonModelNum DL models.\Comment{ Our later empirical results show that even running a total of \numModelsTfAll DL models for diverse tasks in the same setting merely covers a small ratio of possible APIs\lingming{add some precise ratio if possible}\Comment{no fewer than 20\% of the APIs},} Our later empirical results show that they can at most cover \numcradleAPI APIs for \tf, leaving a disproportionately large number of APIs uncovered by such existing techniques.
Secondly, the state-of-the-art model mutation technique proposed by \lemon can be rather limited for generating diverse test inputs for DL APIs.\Comment{the state-of-the-art mutation testing technique for DL libraries is too restricted to effectively generate diverse input for APIs. The technique proposed by \lemon is model-level mutations. We observe that one major drawback of model-level mutation is that it proposes too strong constraints for API invocations.} For example, the intact-layer mutation~\cite{lemon} requires that the output tensor shape of the layer/API to be added/deleted should be identical to its input tensor shape. As a consequence, only a fixed pattern of argument values for a limited set of APIs are explored in model-level mutation, which substantially hinders its bug-finding abilities.
Thirdly, model-level testing can be rather  inefficient.\Comment{ It depends on the external DL datasets to get the input values for tensors, and} The inputs for the original/mutated models are obtained from the external DL datasets, and each of them will need to be completely executed end-to-end to get the final prediction\Comment{\lingming{prediction is better as there can be non-classification tasks?}} results for differential testing, which can consume hours even for a single mutated model. Besides, it requires an additional bug localization procedure to find the specific API invocation causing the inconsistencies between different backends in the original/mutated DL models. During localization, carefully-designed metrics are required to eliminate false positives. The false positives can be due to uncertainty and variances (e.g. floating-point precision loss) in DL APIs, further amplified in the model-level testing scenario.

In this work, we overcome the aforementioned limitations for testing DL libraries via fully automated API-level fuzz testing. Compared with prior model-level DL library testing which resembles \textit{system testing}, API-level testing is more like \textit{unit testing}, which is at a much finer-grained level. The benefit of API-level testing is that it can be a more general and systematic way for testing DL libraries. With API instrumentation, we can get various and diverse input sources from open source to serve the purpose of testing. Moreover, API-level mutation is free of unnecessarily strict constraints on mutation compared with model-level mutation. Besides, API-level mutation neither depends on iterating over external datasets, nor requires complex localization procedures since testing one API at a time does not incur accumulated floating-point precision loss.


\Comment{
\lingming{the next three paragraphs are all about our detailed design and shall be merged into one paragraph}}

One main challenge that we resolve for API-level fuzz testing of DL libraries is fully automated test input generation for DL APIs. The public APIs in DL libraries are mainly exposed in Python, making it difficult to automatically determine the input parameter types due to dynamic typing. To this end, we propose \tech, the first approach to fuzzing DL libraries via mining from actual model and API executions. More specifically, we consider the following sources: 1) code snippets from the library documentation, 2) library developer tests, and 3) DL models in the wild.
\tech records the dynamic information for all the input parameters for each invoked API on the fly while running all the collected code/models. The dynamic information includes the types, values of the arguments, and the shapes of tensors. The traced information can then form a \vs for each API, and an \argdb where values can be shared across arguments of similar APIs during testing. Lastly, \tech leverages the traced information to perform mutation-based fuzzing based on various strategies (i.e., type mutation, random value mutation, and database value mutation), and detects bugs with differential testing and metamorphic testing on different backends. Our initial evaluation of \tech on \pt and \tf shows that \tech can automatically trace valid dynamic information for fuzzing \numTracedAPIs out of all \numHookAPIs considered APIs, while state-of-the-art techniques can at most cover 59\Comment{\chenyuan{should be 59, not inconsistent}} APIs for \tf~\cite{cradle, lemon}. To date, we have submitted \NumAllBugs bug reports (detected by \tech) to developers, with \NumPreviouslyUnknownConfirm already confirmed by developers as previously unknown bugs\Comment{\NumPreviouslyUnknownConfirm confirmed} and \numFixedBugs already fixed to date.

\Comment{\david{Our approach outperforms \lemon \david{by covering\Comment{\APImorecovered more the number of APIs,} \morecoveragelemon more line coverage with \overheadlowerfull lower overhead}}
\david{I don't want to raise concerns of why invoking so many APIs only helps us get such a marginal coverage improvement}}

In summary, our paper makes the following contributions:
\begin{itemize}
    \item \textbf{Dimension.} This paper opens a new dimension for fully automated API-level fuzzing of DL libraries via mining from actual code and model\Comment{ \chenyuan{Maybe "models"?} \yinlin{I think it means "model and code" executions} and code} executions in the wild. 
    \item \textbf{Technique.} We implement a practical API-level DL library fuzzing technique, \tech{}, which leverages three different input sources, including code snippets from library documentation, library developer tests, and DL models in the wild. \tech traces the dynamic API invocation information of all input sources via code instrumentation for fuzz testing. 
    \Comment{input parameters of APIs with code instrumentation to form a \vs for each API, and a \argdb where values can be shared across arguments of similar APIs during mutation testing.} \tech also resolves the test oracle problem with differential testing and metamorphic testing.
    \item \textbf{Study.} Our extensive study on the two most popular DL libraries, \pt and \tf, shows that \tech{} can successfully trace \numTracedAPIs out of \numHookAPIs APIs, and effectively detect \NumAllBugs bugs, with \NumPreviouslyUnknownConfirm already confirmed by developers as previously unknown, and \numFixedBugs already fixed.
\end{itemize}


\Comment{
Deep Learning (DL) has been widely adopted in various domains including image classification, segmentation and natural language processing.

Deep learning systems like Pytorch and Tensorflow have been serving as the best practice for training deep learning models. 

However, like traditional software systems, deep learning frameworks are evolving fast, in order to deliver better performance and programmability.

Meanwhile, hardware like GPU and customized accelerators are also changing rapidly, providing special instructions in order to speed up execution for deep learning operators. Hardware vendors along the way provide the highly optimized libraries for deep learning, making it easier to use.
}

\section{Background}
\label{sec:background}

\subsection{Preliminaries for Deep Learning Libraries}

\begin{figure}[htb]
    \centering
    \includegraphics[keepaspectratio=true,width=\columnwidth]{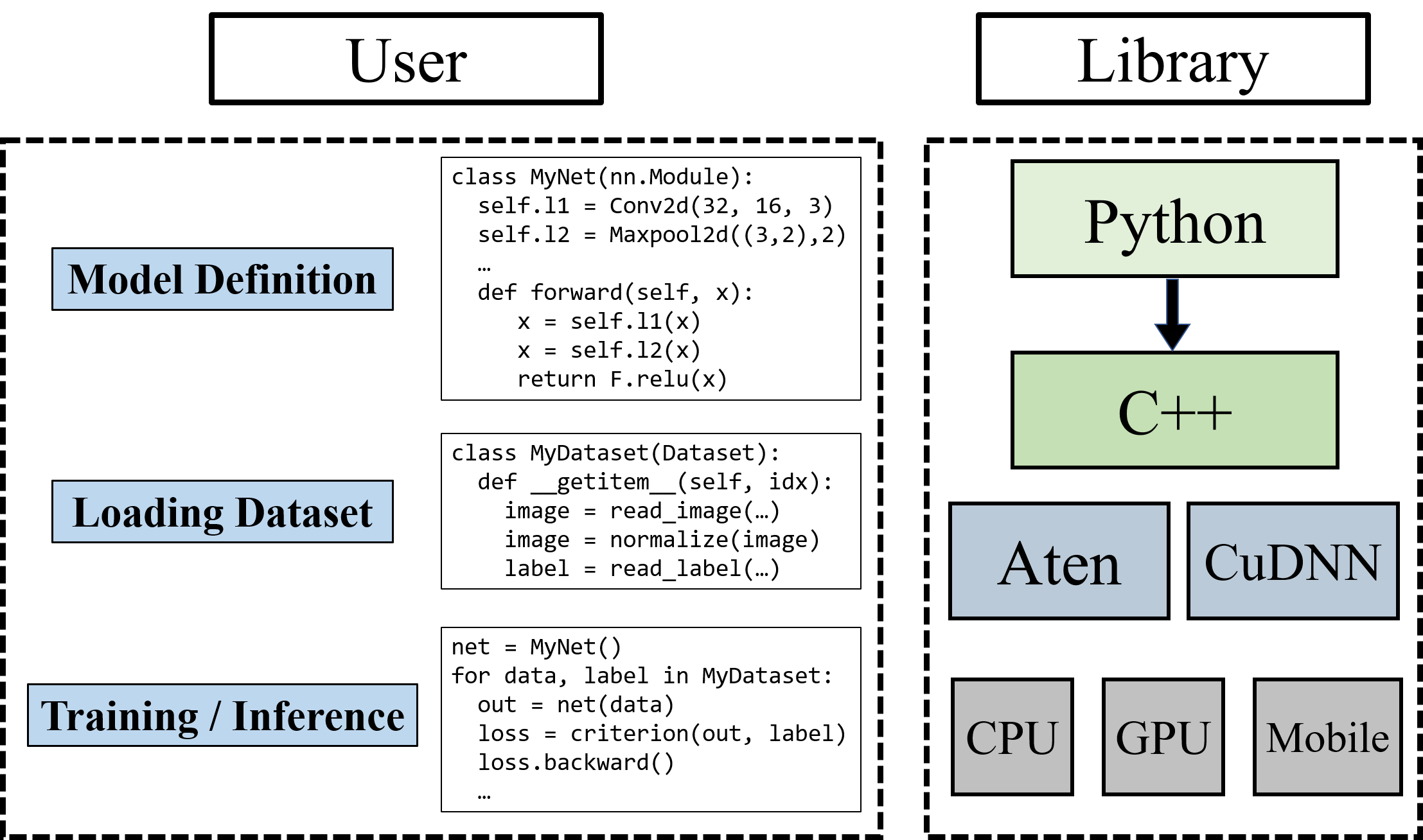}
    \precaptionspace
    \caption{Example DL library (\pt)}
    \label{fig:intro}
\end{figure}

In this section, we will give a brief introduction to the preliminaries of deep learning libraries based on \pt\cite{paszke2019pytorch}.
\Comment{
\lingming{this section talks about various dl lib concepts, we should show a big picture for all of them in a figure similar to the one in lemon}
}
\Comment{
\textbf{Computation graph}\qquad
This figure \david{model representation graph} shows us an example DL model, in the form of \textit{computation graph}. It is a directed acyclic graph (DAG) whose nodes stand for DL operators while edges is used to represent the flow of the data. The data is usually called as \textit{tensor} in DL scenarios, which is a high-dimensional array in essence.
Both frameworks expose the APIs used to construct computation graphs in Python, making it easy for programmers to define a DL model by combining various operators (or layers) to form such a graph, with tensor its input and output. After that, optimizations can be performed on the computation graph in order to achieve better performance. One common practice is operator fusion (or kernel-layer fusion)\cite{niu2021dnnfusion,jia2019taso}, i.e. merging several nodes in the graph into one node to avoid multiple kernel launches. Then the topological order of the graph is used to perform the actual operator execution.\lingming{do we really need to introduce computation graph? maybe better to just start with dl model and mention graph a bit when explaining dl model?  }
}

\parabf{Training and Inference.} As shown on the left-hand side of Figure~\ref{fig:intro}, developers usually leverage DL libraries to support the training and inference tasks on Deep Neural Networks (DNNs). Conceptually, DNNs are composed of \textit{multiple} layers, which are what the adjective ``deep'' in deep learning refers to. In the model definition part of Figure~\ref{fig:intro}, \CodeIn{Conv2d} and \CodeIn{Maxpool2d} are the APIs invoked to add two layers into the example DNN. Then the \CodeIn{forward} function defines how the input data should flow in the defined layers. Before the actual training and inference, the datasets should also be loaded with necessary pre-processing, e.g., {\CodeIn{torchvision.transforms.Normalize}} is a crucial step in data pre-processing, which aims to rescale the values of input and target variables for better performance.

Training is the process for a DNN to learn how to perform a task (with its weights updated along the way). For example, for image classification, by feeding a DNN with known data and corresponding labels, we can obtain a trained DL model. Training a DL model involves iterating over the dataset, defining a loss function {(e.g., \CodeIn{torch.nn.CrossEntropyLoss})} to calculate the difference between the network outputs and its expected outputs (according to the labels), and updating the weights of the DNN via a back-propagation procedure {(i.e., \CodeIn{loss.backward})}. Different from the training phase, inference is the process of using a trained DL model to complete a certain task (with its weights unchanged), e.g., making predictions against previously unseen data based on the trained model.

\Comment{\parabf{DL Model.} Conceptually, DL models are composed of \textit{multiple} layers, which are what the adjective ``deep'' in deep learning refers to. In the model definition part of Figure~\ref{fig:intro}, \CodeIn{Conv2d} and \CodeIn{Maxpool2d} are the APIs invoked to add two layers into the example model. Then the \CodeIn{forward} function defines how the input data should flow in the defined layers.
}
\Comment{Some layers defined in the computation graph have weights automatically attached to it, for instance, when adding the layer \CodeIn{torch.nn.Conv2d} (shown in Figure~\ref{fig:conv-definition}) to the model, the weights corresponding to the layer are automatically attached to the model, which are to be updated during training. Other layers do not have weights, such as activation layers including \CodeIn{torch.nn.ReLU}, defined as $ReLU(x)=max(0,x)$.}

\Comment{\parabf{Dataset.}
Arguably, datasets are crucial to deep learning. For supervised learning tasks, high-quality labeled datasets are needed. Data pre-processing is also an important procedure in deep learning, which involves preparing the data to serve as the input of the DL models. For example, normalization is a crucial step in data pre-processing, which aims to rescale the values of input variables and target variables.}

\Comment{
\lingming{better show some example apis used in training and inference}
}

\Comment{
An inevitable process during training the network is backpropagation, which computes the gradient of the loss function with respect to the weights of the network.
Automatic differentiation is a technique implemented in both \pt  and \tf to automatically calculate the gradient by applying chain rule on the computational graph. For the next iteration of training process, the weights in the DNN are updated with the help of user-defined optimizers. Optimizers (such as stochastic gradient descent) are a set of gradient-based optimization algorithms which can update the gradients in a smart way to accelerate and smoothen the training process.
}

\Comment{
\textbf{Data pre-processing}\qquad
Besides inference and training, data pre-processing is another important procedure in deep learning. It involves preparing the data to serve as the input of the DL model. For example, normalization is a crucial step in data pre-processing, which aims to rescale the values of input variables and target variables. Unscaled input variables often lead to an unstable learning process, and unscaled target variables may result in exploding gradients, making the weight values change dramatically and thus destabilizing the training process. Besides, data augmentation is another pre-processing technique, widely used in image-based deep learning tasks in order to increase the amount and variance of training data. For images, data augmentation strategies like rotation, flipping, cropping are frequently adopted.
In this sense, APIs frequently used during data pre-processing can also be regarded as DL operators, e.g. \CodeIn{torch.transpose}, \CodeIn{torch.flip}, etc.
\Comment{Well in tf.keras they are actually trying to refactor pre-processing operators as layers (for better abstraction and optimization I guess): https:--tensorflow.google.cn-api_docs-python-tf-keras-layers-experimental-preprocessing }
}

\parabf{Abstraction for Hardware.}
\Comment{\lingming{can this one be merged with the next paragraph?}
}
Shown on the right-hand side of Figure~\ref{fig:intro}, DL libraries (such as \pt and \tf) usually provide a unified abstraction for different hardware, which can be easily configured by the end users to perform the actual execution. They usually integrate different backends in DL libraries for flexibility and performance. Take \pt as an example, Aten~\cite{aten} is a backend implemented in C++ serving as a tensor library for hundreds of operations. It has specialized low-level implementations for hardware including both CPUs and GPUs.
Besides Aten, CuDNN~\cite{chetlur2014cudnn} is another backend integrated into \pt, which is a widely-used third-party high-performance library, developed specifically for deep learning tasks on Nvidia GPUs.
Furthermore, as shown in Figure~\ref{fig:intro}, \pt now not only supports general-purpose devices such as CPUs and GPUs, but also allows users to run DL models on mobile devices due to the growing need to execute DL models on edge devices.
\Comment{
\textbf{Integration of high-performance backends}\qquad
Different library backends are integrated in DL libraries. Take \pt as an example, Aten\cite{aten} is a backend implemented in C++ serving as a tensor library for many hundreds of operations. It has specialized low-level implementations for hardware including both CPUs and GPUs.
Besides Aten, CuDNN\cite{chetlur2014cudnn} is another backend integrated into \pt, which is a widely-used third-party high-performance library, developed specialized for Nvidia GPUs.
Generally speaking, these backends implemented in C++ focus on GPU acceleration and fast CPU performance.
For end users, they can freely and easily choose different backends and hardware either in Python code or by setting environment variables in shell before execution.
}
\Comment{
\textbf{Continuous evolution}\qquad
\lingming{this is different from the other dl lib concepts. maybe move this into the challenges for testing dl libs, e.g., dl libs are 1) evolving, and 2) dynamically typed}
While more and more new DL operators and algorithms are being proposed, in the meantime, new data types and hardware with new instructions specially designed for DL workloads are also gaining more and more attention, thus making DL libraries fastly evolving.

From a hardware perspective, devices such as ARM Mali GPU (mobile platform), AI chip NPU\cite{zhao2021akg}, Nvidia A100 GPUs are emerging, providing hardware instruction-level  support (e.g. Tensor Core WMMA instruction) for convolution and matrix operations for acceleration. Data types like FP16 (also known as half-precision), TensorFloat-32 (TF32), BFLOAT16, INT8 (8-bit integer), have been proposed in order to trade a bit of precision for less memory consumption and better performance, and hardware like Nvidia A100 can support them well with sigificant acceleration compared with with the default FP32 (single-precision) data type.

From a software perspective, techniques like mixed precision training\cite{micikevicius2017mixed}, post-training quantization and quantization aware training\cite{han2015deep}, have been proposed to exploit the extra benefits from lower-precision, namely the additional hardware acceleration for specialized data types. 

Implementing the cutting-edge algorithms and techniques while seamlessly supporting various newly-emerging hardware and data types makes developing DL libraries a grand challenge. Prior work on testing DL libraries (\cradle\cite{cradle} and \lemon\cite{lemon}) focuses on Theano, CNTK, MXNet, and \tf. However, Theano is no longer actively maintained by developers since 2017\cite{theano_abandoned}: ``After almost ten years of development, we have the regret to announce that we will put an end to our Theano development.'' Similarly, CNTK's last official release date is in the March of 2019. \pt, \tf and MXNet are constantly evolving in order to keep pace with the latest advances in AI research.
}
\subsection{Fuzzing Deep Learning libraries}
\begin{figure}
    \centering
    \includegraphics[keepaspectratio=true,width=\columnwidth]{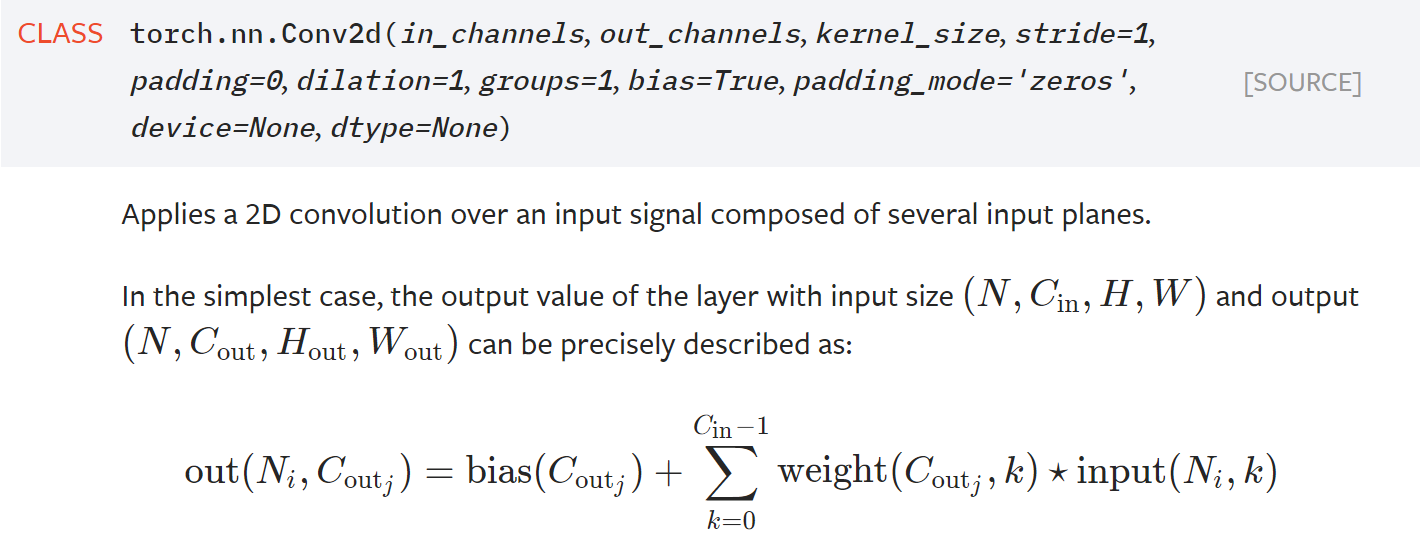}
    \precaptionspace
    \caption{The API definition for \convtwod in \pt}
    \label{fig:conv-definition}
\end{figure}

As shown in the previous subsection, hundreds or even thousands of APIs are implemented in a typical DL library to support various tasks. Therefore, it is almost impossible to manually construct test inputs for each API.
Meanwhile, most public APIs from DL libraries are exposed in Python due to its popularity and simplicity, which makes it extremely challenging to automatically generate test inputs given the API definitions. The reason is that we cannot determine the types of the input parameters statically because Python is a dynamically typed language.\Comment{ Therefore, the main challenge to automatically generate test inputs given the definition of one API is that we cannot determine the types of the input parameters statically because Python is a dynamically typed language.}
\Comment{Python's dynamic typing mechanism means that the type of the variable is determined only during runtime, making it hard to automatically generate test input for APIs.} Take the operator \convtwod from \pt as an example, the definition of which is shown in Figure~\ref{fig:conv-definition}, a snapshot captured from Pytorch official documentation~\cite{conv2d_website_pt}.
From the definition of \convtwod shown Figure~\ref{fig:conv-definition}, we do not know what types of parameters \CodeIn{in\_channels}, \CodeIn{out\_channels}, \CodeIn{kernel\_size} are. Also, one may conclude that parameter \textit{stride} must be an integer\Comment{ \chenyuan{an integer}} (inferred from the default value \CodeIn{stride=1}) and parameter \textit{padding} must also be an integer\Comment{\chenyuan{an integer}} (inferred from the default value \CodeIn{padding=0}). However, this is not the case actually. The documentation below (not included in Figure~\ref{fig:conv-definition} due to space limitations) says that ``stride controls the stride for the cross-correlation, \textit{a single number or a tuple}'' and ``padding controls the amount of padding applied to the input. It can be \textit{either a string {‘valid’, ‘same’} or a tuple of ints} giving the amount of implicit padding applied on both sides''. In fact, the parameters \CodeIn{kernel\_size}, \CodeIn{stride}, \CodeIn{padding}, \CodeIn{dilation} can be either a single \textit{int} or a \textit{tuple of two ints}, and \CodeIn{padding} can even be a \textit{string} besides the two types mentioned above. Therefore, there can exist multiple valid types for a specific argument, and the valid types of arguments cannot be easily inferred from the definition.

\Comment{
A recent work on precision testing for DL operators \predoo~\cite{zhang2021predoo} tries to manually maintain a seed pool for the test input generation (including tensors' shape, and necessary parameters required by each operator), and it tests only \predooOperatorNum selected operators from \tf. Such a methodology suffers from scalability, which cannot be applied to many hundreds of operators in DL libraries.
}
\Comment{
\lingming{we should also talk about craddle (and other dl lib testing work if any), and reduce the text for lemon (by around 50\%; please do not directly remove texts, only comment them out so that we may later move them to other sections)}
}

Due to the above challenge of test generation for DL APIs, \cradle~\cite{cradle} proposes to directly leverage existing DL models to test DL libraries. The insight of \cradle is to check the cross-implementation inconsistencies when executing the same DL models on different backends to detect DL library bugs. It uses \cradleModelNum models and \cradleDatasetNum datasets. After detecting inconsistencies when executing models between different backends by feeding the input from datasets, a confirmation procedure to identify bug-revealing inconsistencies and a localization procedure to precisely localize the source of the inconsistencies have to be launched. In such model-level testing, where inconsistencies can be either due to real bugs or accumulated floating point precision loss
 propagated through the execution of multiple APIs, carefully designed metrics are needed to distinguish real bugs from false positives. 
Furthermore, such model-level testing technique only covers a limited range of APIs in DL libraries, e.g., all models used by \cradle only cover \numcradleAPI APIs for \tf. \Comment{give exact ratio after finishing experiments for tf1.14}\Comment{, and it may suffer from unwanted inconsistencies which are false positives}


Based on \cradle, \lemon~\cite{lemon} advances testing DL libraries by proposing model-level mutation. A set of model-level mutation rules are designed to generate mutated models, with the goal of invoking more library code. Model-level mutation is composed of intact-layer mutation and inner-layer mutation.
The intact-layer mutation rules pose very strict constraints on the set of APIs to be mutated and the arguments passed to them. As stated in the \lemon paper~\cite{lemon}, one explicit constraint for intact-layer mutation is that the output shape of the API to be inserted and the input shape of it must be identical. As a result, only a limited set of APIs with fixed parameters can used for mutation in order to meet such constraints, which substantially hinders \lemon's ability in bug-finding. Moreover, selecting such APIs with specific arguments for layer-level mutation requires expert knowledge of the input-output relation of each API. For example, only a limited range of APIs (e.g., convolution, linear and pooling) with fixed parameters can be added or deleted during model-level mutation. According to our later study, \lemon's various mutation rules can only help cover 5 more APIs in total for all the studied models. \Comment{\lingming{add some texts showing model-level mutation barely increases API and code cov}}
\Comment{
Take the operator \convtwod as an example, with its API definition from \pt shown in Figure~\ref{fig:conv-definition}. The input to \convtwod layer should be a 4-dimensional tensor with $(N, C_{in}, H_{in}, W_{in})$ as its shape and its output is also a 4-dimensional tensor with $(N, C_{out}, H_{out}, W_{out})$ as its shape. The input-output relation for \convtwod in \pt is the following formula where \textit{pad}, \textit{dil}, \textit{ker} stands for \CodeIn{padding}, \CodeIn{dilation}, \CodeIn{kernel\_size} respectively, which are passed in (together with \CodeIn{out\_channels}) as API parameters:
\[
C_{out} = out\_channels 
\]
\[
H_{out} = \lfloor \frac{H_{in} + 2 \times pad[0] - dil[0] \times (ker[0] - 1) - 1}{stride[0]} + 1 \rfloor
\]
\[
W_{out} = \lfloor \frac{W_{in} + 2 \times pad[1] - dil[1] \times (ker[1] - 1) -1}{stride[1]} + 1 \rfloor
\]

In \keras, the API for \convtwod is similar. One slight difference is that the input and output tensor's shapes are $(N, H_{in}, W_{in}, C_{in})$ and $(N, H_{out}, W_{out}, C_{out})$ respectively as default (also known as the ``channels\_last mode'').
In the official open source implementation~\cite{lemon_website} of \lemon \Comment{shown in Figure~\ref{fig:lemoncode}, Line~\ref{line:kerasconv2d1} and Line~\ref{line:kerasconv2d2} constructs a \convtwod layer to be added to the original model as one of its intact-layer mutation rule called \textit{Layer Addition}.
The first parameter \CodeIn{(input\_shape[-1]} (namely $C_{in}$ as the last value of $(N, H_{in}, W_{in}, C_{in})$) of the API \CodeIn{keras.layers.Conv2D} is passed in to determine the output channels $out\_channels$, which means that $C_{out} = C_{in}$.}
The following parameters (\CodeIn{3} and \CodeIn{strides=(1,1)}) indicate that $ker[0] = ker[1] = 3$. Combining the above analysis with $stride[0] = stride[1] = 1,$ \quad $pad[0] = pad[1] = 1$ (default values of the API), we can safely conclude that $H_{out} = H_{in}, W_{out} = W_{in}, C_{out} = C_{in}$ using the formula.
}

\Comment{
In order to check whether the \convtwod layer can be added in the original model, some pre-conditions are checked in \lemon. More specifically, the input tensor's shape is checked from Line~\ref{line:inputlegalstart} to Line~\ref{line:inputlegalend} in Figure~\ref{fig:lemoncode}, namely ensuring that the dimension of the input tensor is exactly 4 (specified by the API), the batch dimension $N$ (\CodeIn{input\_shape[0]}) is not a specific concrete value, and $W_{in}$ and $C_{in}$ are greater than 3 (with no strong rationale).

\begin{figure}[t]
	\begin{lstlisting}[basicstyle=\ttfamily\footnotesize,escapeinside={(*@}{@*)},columns=fixed,xleftmargin=3.5ex,numbers=left,language=Python]
def conv2d(input_shape):
    import keras
    layer = keras.layers.Conv2D(input_shape[-1], 3,\(*@\label{line:kerasconv2d1}@*)
                strides=(1,1), padding='same')(*@\label{line:kerasconv2d2}@*)
    layer.name += '_insert'
    return layer

def conv2d_input_legal(input_shape):
    input_shape = input_shape.as_list()
    return len(input_shape) == 4 \    (*@\label{line:inputlegalstart}@*)
        and input_shape[0] is None \
        and input_shape[1] is not None \
        and input_shape[1] >= 3 \
        and input_shape[2] is not None \
        and input_shape[2] >= 3(*@\label{line:inputlegalend}@*)
    \end{lstlisting}
	\caption{Mutation rule for \convtwod from \lemon}
	\label{fig:lemoncode}
\end{figure}
}

\Comment{
Such strong constraints on the API invocations, which not only substantially hinder \lemon's ability to detect more bugs, but also require huge manual efforts and domain-specific knowledge, are required to exist in \lemon due to the following two reasons: one is the internal constraints imposed by APIs, and the other is the technique of model-level testing proposed by \lemon. The internal constraints are easy to understand. For example, constraints of the \convtwod API in \pt shown in Figure~\ref{fig:conv-definition} include reasonable values with the correct types to be passed in (e.g. \CodeIn{stride} is an integer\Comment{ \chenyuan{an integer}} or a tuple of two ints), and the correct shape of input tensor (it must receive a 4-dimensional tensor). Actually the technique of model-level testing is to blame for unnecessarily strong constraints for API invocation. The mutation rules at the model level are designed to generate more diverse models, in other words, to construct different computation graphs. Adding one node into the graph while not breaking other nodes' execution has to make sure that the input tensor and output tensor of the API to be inserted or removed have exactly the same shape. This is ensured in \lemon by either invoking APIs by fixed parameters with strong constraints, or delibrately selecting such elementwise APIs (activation functions including \CodeIn{relu, tanh, sigmoid, leakyrelu}) with domain-specific knowledge, otherwise the whole model is not valid at all and cannot be successfully executed.
}

\Comment{
Furthermore, model-level testing is not ideal because of it requires additional efforts in designing metrics due to randomness. For example, the metric \textit{D\_MAD}\cite{cradle} is proposed to measure the degree of inconsistencies (or distance) between the final prediction vector and the ground-truth vector. Also, layer localization metric is proposed to localize the root cause layer for such inconsistencies. However, we argue that our finer-grained API-level testing relieves us from such burdens, because no accumulated randomness is introduced and we do not observe any difficulty in reproducing the computation results of each API.
}

\Comment{
Last but not least, the inner-layer mutation rules that \lemon proposes are largely irrelevant to bug-finding in DL libraries. For example, the \textit{Gaussian Fuzzing} strategy that \lemon uses aims to add noise to the weights of a layer, and the \textit{Neuron Effect Block} strategy sets the weights of a layer to be 0. \lingming{this paragraph should be largely reduced or removed. as i said before, whether the lemon mutators were used in prior work solving other problems or not does not matter; it is novel as long as it is applied for a different task here, i.e., detecting dl lib bugs. }These strategies can also be found in DeepMutation\cite{ma2018deepmutation} where it is the first time to propose a set of mutations strategies including Gaussian Fuzzing and Neuron Effect. These strategies are designed by DeepMutation to measure the quality of test data by injecting faults  (namely deliberately changing the weights) into a trained model. The only influence of applying these mutations rules is changing the values of weights of trained model, which is a perfect fit for measuring the quality of test data (as the research goal of DeepMutation\cite{ma2018deepmutation}), but has little effect on fuzzing DL libraries because it neither invokes more APIs nor changes the function parameters. Another evidence is that we cannot easily observe any explicit correlation between the bugs detected by \lemon and the inner-layer mutation rules that \lemon proposes.
}



\section{Approach}
\begin{figure*}[htb]
    \centering
    \includegraphics[keepaspectratio=true,width=\textwidth]{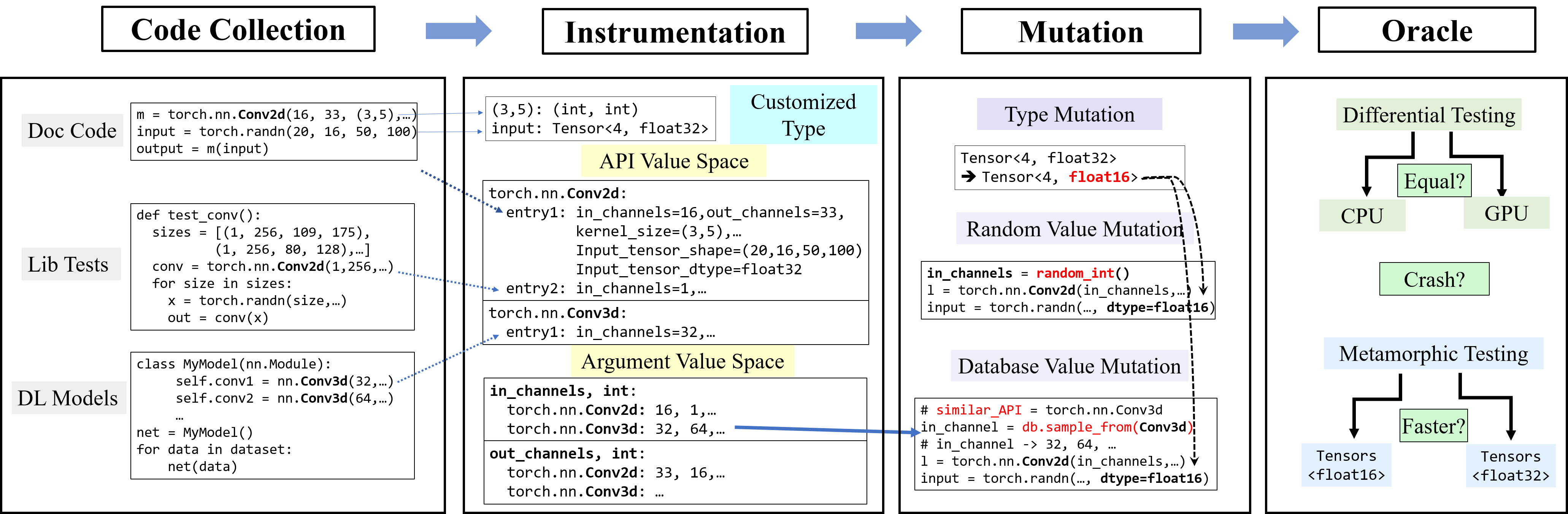}
    \caption{\tech overview}
    \label{fig:overview}
\end{figure*}

Figure~\ref{fig:overview} shows the overview of our approach, \tech, which is mainly composed of four different stages.
The first stage is code collection (Section~\ref{subsec:codecollection}). As shown in the figure, \tech obtains code from three different sources: 1) code snippets from library documentation, 2) library developer tests, and 3) various DL models in the wild, all of which can be obtained from open source automatically.
The second stage is dynamic tracing with instrumentation (Section~\ref{subsec:instrumentation}). \tech first hooks the invocation of APIs, and then executes the code collected in the first stage to trace various dynamic execution information for each API invocation, including value and type information for all parameters of all executed APIs. As a result of this stage, \tech constructs the type space, API value space, and argument value space for the later fuzzing stage.
The third stage is mutation-based fuzzing (Section~\ref{subsec:mutation}). Basically, \tech effectively generates mutants for the test inputs (i.e., the argument lists) used to invoke the targeted APIs, based on the traced information collected in the second stage. The mutation strategies are composed of type mutation, random value mutation, and database value mutation.
The last stage is running all the generated tests with oracles (Section~\ref{subsec:testoracle}). \tech resolves the test oracle problem by differential testing and metamorphic testing on different DL library backends and hardware.
\tech is able to detect various types of bugs, including wrong-computation bugs, crash bugs, and performance bugs for DL libraries.

\subsection{Code Collection}
\label{subsec:codecollection}
\begin{figure}[htb]
    \centering
    \includegraphics[keepaspectratio=true,width=\columnwidth]{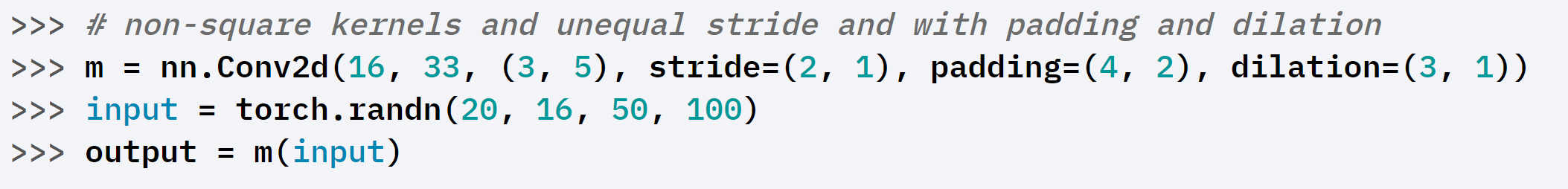}
    \precaptionspace
    \caption{Example Code for \convtwod from \pt's Documentation}
    \label{fig:conv-code}
\end{figure}

\tech is a general approach and can work with dynamic API information traced from various types of code executions. In this paper, we mainly explore code collection from the following sources:

\parabf{Code Snippets from Library Documentation.} In order to help users better understand the usage of APIs, almost all DL libraries will provide detailed documentations on how to invoke the APIs. 
Usually, detailed specifications written in natural languages are presented to show the usage of each parameter of each API in detail. Meanwhile, to better help the developers, such natural-language-based specifications are also often accompanied by code snippets for better illustrations.
To illustrate, an example code snippet for invoking the \convtwod API within \pt is shown in Figure~\ref{fig:conv-code}. Of course, it is worth noting that not all APIs have example code and example code cannot enumerate all possible parameter values\Comment{ exploring the parameters of each API to the users}. Therefore, it is also important to consider other sources.

\parabf{Library Developer Tests.} Software testing has become the most widely adopted way for quality assurance of software systems in practice. As a result, DL library developers also write/generate a large number of tests to ensure the reliability and correctness of DL libraries. For example, the popular \tf and \pt DL libraries have \numtestpt and \numtesttf tests for testing the Python APIs, respectively.
\Comment{let's simply say we use all the tests for testing the Python APIs (without mentioning anything about the C++ tests; i can help check later)}
We simply run all such Python tests as they dominate DL library testing and this work targets Python API fuzzing.

\parabf{DL Models in the Wild.} Popular DL libraries have been widely used for training and deploying DL models in the wild. Therefore, we can easily collect a large number of models for a number of diverse tasks, each of which will cover various APIs during model training and inference. More specifically, from popular repositories in Github, we obtain \numModelsPtAll models for \pt, and \numModelsTfAll models for \tf. These models are diverse in that they cover various tasks such as image classification,\Comment{ object detection,} natural language processing, reinforcement learning, autonomous driving, etc. The detailed information about the leveraged models can be found in our repository~\cite{freefuzzrepo}.\Comment{Section~\ref{tab: modelcompare}.}


\subsection{Instrumentation}
\label{subsec:instrumentation}
In this phase, \tech performs code instrumentation to collect various dynamic execution for test-input generation. We first get the list of \Comment{all possible }Python APIs to be instrumented from the official documentations of the studied DL libraries in this work, i.e., \pt and \tf. More specifically, we hook the invocation of the list of \numHookAPIsPt APIs from \pt and \numHookAPIsTftwo APIs from \tf for dynamic tracing.
\Comment{\lingming{instead of saying we ignore what APIs, it is better to say we include all APIs satisfying some constraints, e.g., public APIs for what computations!}
\david{More specifically, we include all the necessary APIs for building neural networks and tensor computations (or interchangebly,  data processing, which seems better?). Also, simply `tensor computations' without mentioning building neural networks also makes sense to me.}}
The list includes all the necessary APIs for training and inference of neural networks as well as performing tensor computations.
\tech{} is able to collect\Comment{ various} dynamic information for each API invoked by all the three sources of code/model executions, including the type and value for each parameter.
\Comment{\lingming{are tensors are also parameter?}\david{it depends, let me avoid confusion here.}}
No matter how the APIs are invoked (e.g., executed in code snippets, tests, or models), the corresponding runtime information of the arguments is recorded to form the following type/value spaces for fuzzing:\Comment{ the API value space and a database of argument value space.}

\begin{figure}[htb]
    \centering
    \includegraphics[keepaspectratio=true,width=\columnwidth]{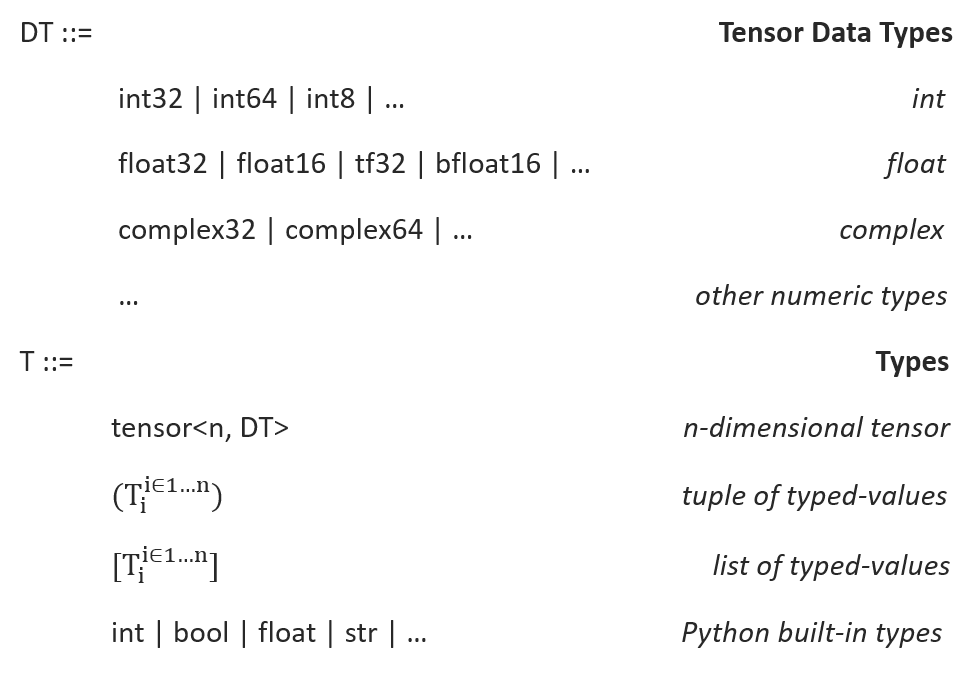}
    \precaptionspace
    \caption{Customized Type Monitoring System \fuzztype}
    \label{fig:type}
\end{figure}

\parabf{Customized Type Space.}
\tech constructs our customized type monitoring system \fuzztype for API parameters by dynamically recording the type of each parameter during API invocation. 
Compared with Python's original type system, the customized type system is at a finer-grained level, which can better guide the next mutation phase for fuzzing.
In Python's dynamic type system, the type of parameter \CodeIn{stride=(2,1)} (shown in Figure~\ref{fig:conv-code}) can be calculated by running \CodeIn{type((2,1))}. This will return \CodeIn{<class 'tuple'>}\Comment{\lingming{just curious: is there really no way to get integer tuple info in Python runtime? }\david{we manage to do it based on Python's type}}, which does not encode all the necessary useful information for fuzzing because we only know that it is a tuple.
In our type monitoring system \fuzztype, we collect such information at a finer-grained level: \fuzztype\CodeIn{((2,1))} returns \CodeIn{(int, int)} (a tuple of two integers). Similarly, for tensors, executing \CodeIn{type(torch.randn(20,16,50,100))} simply returns \CodeIn{<class 'torch.Tensor'>} in Python's type system while we can obtain finer-grained type \CodeIn{Tensor<4,float32>} (a 4-dimensional tensor with \CodeIn{torch.float32} as its data type) by running \fuzztype\CodeIn{(torch.randn(20,16,50,100))}.
Our customized type monitoring system used to guide API-level fuzzing of DL libraries is formally defined in Figure~\ref{fig:type}.
\Comment{
\begin{small}
\begin{flalign*}
& DT ::= \tag*{\textbf{Tensor Data Types}}\\
& \qquad\qquad int32 \smallspace | \smallspace int64  \smallspace | \smallspace int8 \smallspace | \smallspace ...\tag*{\textit{int}}\\
& \qquad\qquad float32 \smallspace |   
\smallspace float16 \smallspace | \smallspace tf32 \smallspace | \smallspace bfloat16 \smallspace | \smallspace ...\tag*{\textit{float}}\\
& \qquad\qquad complex32 \smallspace | \smallspace complex64 \smallspace | \smallspace ... \tag*{\textit{complex}}\\
& \qquad\qquad ... \tag*{\textit{other numeric types}}\\
& T ::= \tag*{\textbf{Types:}}\\
& \qquad\qquad tensor<n, DT> \tag*{\textit{n-dimensional tensors}}\\
& \qquad\qquad (T_i^{i\in 1...n}) \tag*{\textit{tuples of typed-values}}\\
& \qquad\qquad  [T_i^{i\in 1...n}] \tag*{\textit{lists of typed-values}}\\
& \qquad\qquad int \tag*{\textit{integers}}\\
& \qquad\qquad bool \tag*{\textit{booleans}}\\
& \qquad\qquad float \tag*{\textit{floating point numbers}}\\
& \qquad \qquad str \tag*{\textit{strings}}\\
& \qquad\qquad ... \tag*{\textit{other Python types}}\\
\end{flalign*}
\end{small}
}

\begin{table*}\centering
\caption{Type Mutation}\label{tab: typemutation}
\scalebox{0.9}{
\begin{tabular}{c|c|c}
\hline
Mutation Strategies &$T_1$ &$T_2$ \\\cline{1-3}
\tensordim & $tensor\lbracket n_1,DT\rbracket$ & $tensor\lbracket n_2,DT\rbracket~(| n_2 - n_1 | > 0)$ \\
\tensordtype & $tensor\lbracket n,DT_1\rbracket$ & $tensor\lbracket n,DT_2\rbracket~(DT_2 \neq DT_1)$\\
\typetransform & $T_1 = int | bool | float | str$ & $T_2 ~(T_2 \neq T_1)$ \\
\tuplemutation & $(T_i^{i \in {1...n}})$ & $(type\_mutate(T_i)^{i \in {1...n}})$\\
\listmutation & $[T_i^{i \in {1...n}}]$ & $[type\_mutate(T_i)^{i \in {1...n}}]$\\
\hline
\end{tabular}}
\end{table*}

\begin{table*}\centering
\caption{Value Mutation}\label{tab: valuemutation}
\scalebox{0.9}{
\begin{tabular}{c|c|c}
\hline
Mutation Strategies &$T$ &$V$ \\\cline{1-3}
\randomtensorshape & $tensor\lbracket n,DT\rbracket$ & $tensor(shape = [randint()^n], datatype=DT)$\\
\randomtensorvalue & $v: tensor\lbracket n,DT\rbracket$ & $tensor(shape=v.shape, datatype=DT).rand()$\\
\randomprimitive & $int|float|bool|str$ & $rand(int|float|bool|str)$\\
\randomtuple & $(T_i^{i \in {1...n}})$ & $(value\_mutate(T_i)^{i \in {1...n}})$\\
\randomlist & $[T_i^{i \in {1...n}}]$ & $[value\_mutate(T_i)^{i \in {1...n}}]$\\
\hline
\dbtensorshape & $tensor\lbracket n,DT\rbracket$ & $pick\_shape(database, tensor\lbracket n,DT\rbracket)$\\
\dbtensorvalue & $tensor\lbracket n,DT\rbracket$ & $pick\_value(database, tensor\lbracket n, DT\rbracket)$\\
\dbprimitive & $int|float|str$ & $pick(database, int|float|str, argname)$\\
\dbtuple & $(T_i^{i \in {1...n}})$ & $pick(database, (T_1, T_2,...,T_n), argname)$\\
\dblist & $[T_i^{i \in {1...n}}]$ & $pick(database, [T_1, T_2,...,T_n], argname)$\\
\hline
\end{tabular}}
\end{table*}

Note that type inference for dynamically typed languages (such as Ruby and JavaScript) via dynamic program tracing has been explored in the literature for traditional applications~\cite{an2011dynamic,pradel2015typedevil,andreasen2016trace}. In this work, we further extend such techniques for fuzzing deep learning libraries.\Comment{\lingming{we need to highlight our differences with existing work here, such as tracing tensor shapes, etc.; also we talked about moving the above type monitoring system into figs or tables to avoid the huge space in the last page.} 
\david{\fuzztype augments the Python built-in types by tracing tensors' shape and data types. Also, \fuzztype keeps track of all the types of elements in collections (i.e., tuples and lists) of heterogeneous objects. \fuzztype guides the type-aware mutations, which are discussed in Section~\ref{subsec:mutation}.}}
Different from prior work, \tech{} collects dynamic traces from various sources, including developer tests, code snippets documents, and DL models in the wild; also, \tech{} augments the Python built-in types to trace and mutate tensor shapes and heterogeneous data types.\Comment{\lingming{check if we can further claim that we are the first to use such dynamic type tracing approach for fuzzing} \david{TypeDevil~\cite{pradel2015typedevil} uses dynamic tracing for type inconsistency bug detection. ACSAC'18 (not cited) `TIFF: Using Input Type Inference To Improve Fuzzing' seems to be earlier than us.}
}

\Comment{\david{Type inference from dynamic executions has been explored before for other applications~\cite{an2011dynamic,pradel2015typedevil,andreasen2016trace}, but not specifically for fuzz testing.}}

\parabf{API Value Space.}
\tech constructs the \vs of each API from the concrete values passed into the API during dynamic tracing. One entry in the \apivs stands for one API invocation with its corresponding list of concrete arguments, which is later used in our mutation phase as the starting point to generate more mutants/tests.
\Comment{\david{I changed the example to match overview figure}}
Take Figure~\ref{fig:overview} as an example, \CodeIn{entry1} is added to the \vs of the API \CodeIn{torch.nn.Conv2d} after executing the documentation code in the code collection phase. More specifically, \CodeIn{in\_channels=16, out\_channels=33, kernel\_size=(3,5)} together with some other values (not shown in Figure~\ref{fig:overview} due to limited space) are recorded in \CodeIn{entry1}.  
\Comment{Take Figure~\ref{fig:conv-code} as an example, the value space of the API \CodeIn{torch.nn.Conv2d} is expanded after execution by recording \CodeIn{in\_channels=16, out\_channels=33, kernel\_size=(3,5), stride=(2,1), padding=(4,2), dilation=(3,1)}, and there are also arguments which are passed in using default values, namely \CodeIn{groups=1, bias=True, padding\_mode=`zeros'}, etc.
}
The return value of \CodeIn{nn.Conv2d} is a callable object, and it expects a tensor as its input, which is initialized as \CodeIn{input=torch.randn(20,16,50,100)}, indicating that \CodeIn{input} is a four-dimensional tensor with \CodeIn{(20,16,50,100)} as its shape and the values are randomly initialized. Note that we also record the corresponding shape and data type information for tensors, i.e., \CodeIn{Input\_tensor\_shape=(20,16,50,100), Input\_tensor\_type=float32}. All the function arguments mentioned above constitute \textit{one entry} in the \vs for \CodeIn{nn.Conv2d}. Each invocation can add a new entry into the value space of the invoked API.
\Comment{
\lingming{how did you control to not add duplicated entries, do you check the entry existence everytime? please show more details; also you used db for the next space, what did we use to store the data for the first two spaces?}
\david{I removed the text. Duplicated entries are removed when doing VS analysis for RQ1 (input breakdown), not during execution on the fly.}}

\parabf{Argument Value Space.}\Comment{\lingming{if the diff is only this one focuses on both values and types for arguments while the first two focus on the api level, the names need to be changed, e.g., how about API Type Space, API Value Space, and Argument Space (or Argument DB)?} \david{How about argument value space?}
\lingming{why only this one is in database (e.g., provide the benefits for doing so)? better also explain why the first two spaces are not stored in db. please also clearly describe the column names for the db, and some example data row. also make sure that your designed column names match the interface used in our db-mutation algorithm.}
\david{I have redrawn Figure~\ref{fig:overview} and rewritten text, to be clearer}}
\Comment{
Different from API value space where one entry stands for one complete argument list for one API invocation, argument value space is a database that stores the mapping between the each argument name and all their recorded values and types for each API.
}
\Comment{\lingming{in a DB, argument names should also be column names. every row is a data item. also, one row cannot hold multiple values. that is, every row will be one value of one type for one argument from one api.}
\lingming{missing type info here; there should be four columns argument name, value, and type as well api name?}
\david{it is not a rigorous database actually. Let me change the name to \argdb. I redraw overview figure again.}}
As shown in Figure~\ref{fig:overview}, the \argdb is composed of different argument names and types (e.g. \CodeIn{in\_channels} of type \CodeIn{int}\Comment{ and \CodeIn{out\_channels} of type \CodeIn{int}}), together with their values recorded when invoking different APIs.
For example, for the argument \CodeIn{in\_channels} of the API \CodeIn{torch.nn.Conv2d}, the values recorded include \CodeIn{16,1}, etc. The \argdb is constructed based on the information collected in the \apivs to speed up the queries in our database value mutation strategy discussed later. More specifically, \argdb aggregates values from different APIs based on argument names.
\Comment{
For example, after executing example code of the documentation code and library tests in Figure~\ref{fig:overview}, the entry\lingming{i assume 16 and 1 are two executions, so entries rather than entry?} of \CodeIn{in\_channels} for \CodeIn{torch.nn.Conv2d} will be \CodeIn{16, 1} (and the types are \CodeIn{int}).
}
The \argdb is formed based on the idea that values for an argument of one API can serve as potentially reasonable values for the argument of other similar APIs. For example, \CodeIn{torch.nn.Conv2d} and \CodeIn{torch.nn.Conv3d} can be considered similar.
The API definition of \convthreed is \CodeIn{torch.nn.Conv3d(in\_channels, out\_channels, kernel\_size, stride=1, padding=0, dilation=1, groups=1, bias=True, padding\_mode=`zeros', device=None, dtype=None)}, and many parameters share the same names as \CodeIn{torch.nn.Conv2d} (shown in Figure~\ref{fig:conv-definition}). The construction of the argument value space is useful for the database value mutation to be introduced in the next section.

\subsection{Mutation\Comment{\lingming{no guidance now?}\david{you are right}}}
\label{subsec:mutation}

In this phase, \tech applies various mutation rules to mutate the argument values traced in the second phase to generate more tests for fuzzing DL libraries more thoroughly.

\parabf{Mutation Rules.}
The mutation rules for \tech{} are composed of two parts: \textit{type mutation} and \textit{value mutation}, shown in Tables~\ref{tab: typemutation} and \ref{tab: valuemutation}, respectively.
Type mutation strategies include \textit{\tensordim} that mutates $n_1$-dimensional tensors to $n_2$-dimensional tensors, \textit{\tensordtype} that mutates the data types of tensors without changing their shapes, \textit{\typetransform} that mutates one primitive type into another, as well as \textit{\tuplemutation} and \textit{\listmutation} that mutate the types of elements in collections of heterogeneous objects.

Value mutation strategies are divided into two classes: one is random value mutation, and the other is database value mutation.
Random value mutation strategies include \textit{\randomtensorshape} (using random integers as shapes to mutate $n$-dimensional tensors\Comment{while maintaining original data types}), \textit{\randomtensorvalue} (using random values to initialize tensors\Comment{while maintaining original data types}), \textit{\randomprimitive}, \textit{\randomtuple} and \textit{\randomlist}. Database mutation strategies include \textit{\dbtensorshape} and \textit{\dbtensorvalue}, which randomly pick the according shapes or values from database of \argdb, together with \textit{\dbprimitive}, \textit{\dbtuple}, and \textit{\dblist}, which randomly pick the corresponding entries from the \argdb based on the argument names and types. Note that all the mutation rules are type-aware, i.e., they are applied according to the types.

\parabf{Algorithm.}
Shown in Algorithm~\ref{alg:mutation}, the input to our fuzzing algorithm is the API to be mutated, entries in the API value space \CodeIn{VS}, 
and the database of argument value space \CodeIn{DB}. Of course, we also need to define the mutation rules as described above.
In each iteration, the algorithm always samples the next entry from the API value space \CodeIn{VS[API]} to start the mutation process (Line~\ref{lst:line:selectnext}). \tech then computes the number of arguments \CodeIn{argNum} in the entry (Line~\ref{lst:line:argnum}), and randomly selects an integer\Comment{ \chenyuan{an integer}} between 1 and \CodeIn{argNum} as the number of arguments to be mutated, i.e., \CodeIn{numMutation} (Line~\ref{lst:line:numMutation}). Then, \tech starts an inner loop to mutate \CodeIn{numMutation} arguments to generate a new test.
The arguments are mutated one by one via randomly selecting a random argument index \CodeIn{argIndex} (Line~\ref{lst:line:argindex}). After determining the argument to be mutated each time, \tech gets the type of it using our customized type system \fuzztype, the argument name \CodeIn{argName}, and the argument value \CodeIn{argValue} (Lines~\ref{lst:line:type}, ~\ref{lst:line:argname} and~\ref{lst:line:argvalue}). The type mutation will be performed nondeterministically -- if it is enabled, \tech will mutate the argument type according to our type mutation strategies (Line~\ref{lst:line:typerule}). \CodeIn{selectRandOverDB} is another random function called to determine whether to perform random value mutation (Line~\ref{lst:line:randvalue}) or database value mutation (Line~\ref{lst:line:dbvalue}) according to the corresponding mutation rules. After mutating \CodeIn{numMutation} arguments for \CodeIn{entry}, \tech generates a new test, which will be executed for testing the API (Line~\ref{lst:line:run}). Then, the main loop will continue to generate the next test until the termination criterion is met, e.g., generating a specific number of new tests\Comment{ or reaching a timeout}.

\newcommand{\lev}{Levenshtein}
\newcommand{\apione}{API_1}
\newcommand{\apitwo}{API_2}
\newcommand{\apii}{API_i}
\newcommand{\apij}{API_j}

\newcommand{\simFunc}{Sim}
\newcommand{\probFunc}{Prob}
\newcommand{\simVar}{sim}
\newcommand{\probVar}{prob}

We next discuss function \CodeIn{$ValueRule_{db}$} in more detail to explain the process for mutating the value of an argument for a specific API based on the \argdb. Shown in the algorithm, the function takes the API name \CodeIn{API}, the type of argument \CodeIn{argType}, the name of the argument \CodeIn{argName}, and the database \CodeIn{DB}, as input parameters.
It then queries the database to collect all the APIs which share the same argument name and type as the current API under test (Line~\ref{lst:line:apiquery}). Next, \tech computes the text similarities between the current API under test and each of the returned APIs based on the Levenshtein Distance~\cite{levenshteindistance} between API definitions (Line~\ref{lst:line:simapi}). Take the query \CodeIn{$ValueRule_{db}$(torch.nn.MaxPool2d, [int, int], `dilation', DB)} as an example, the text similarity is computed using API definitions of those containing the same argument name (\CodeIn{'dilation'}) and the type (tuple of two integers). More specifically, the similarity between the current API under test and $\apii$, one of the returned APIs, can be computed by the following formula: 
$$\simFunc(\apii, API)=1 - \frac{\lev(\apii, API) } {Max(Len(\apii), Len(API))}$$ where function $\lev$ computes Levenshtein Distance between the two strings representing $\apii$ and $API$, and it is divided by the maximum length of the two strings. The whole formula computes the text similarity of the two API definitions. For our example, the result shows that the definition of \CodeIn{torch.nn.Conv2d} has the highest text similarity with the target API \CodeIn{torch.nn.MaxPool2d(kernel\_size, stride=None, padding=0, dilation=1, return\_indices=False,}\\ \CodeIn{ceil\_mode=False)}\Comment{, compared with two other candidates (\CodeIn{torch.nn.functional.conv\_transpose2d} and \CodeIn{torch.nn.functional.conv2d})}. Then we normalize the text similarities to transform them into probabilities (summing up to 1) for selecting similar APIs (Line~\ref{lst:line:softmax}). The basic idea is that APIs with higher similarity scores should get higher probabilities to be selected. \tech does this by performing the Softmax computation~\cite{softmax}\Comment{\lingming{add citation to softmax}}: 
$$\probFunc(API_i)=\frac{e ^ {\simFunc(API_i, API)}}{\Sigma_{j=1}^{m}e ^ {\simFunc(API_j, API)}}$$
where $m$ denotes the number of APIs sharing the same argument name and type as the current API under test. After sampling a random API according to the probabilities (Line~\ref{lst:line:apiselected}), the values are then randomly sampled from the values recorded for the API (Line~\ref{lst:line:valueselected}).\Comment{The values of ``stride" recorded for \CodeIn{Conv2d} include \CodeIn{[1, 1], [3, 1], ...}, which has a probability of 0.56 to be sampled for the usage of \CodeIn{MaxPool2d}. The values in \CodeIn{torch.nn.functional.conv\_transpose2d} and \CodeIn{torch.nn.functional.conv2d} have probabilities of 0.21 and 0.23 to be sampled respectively.} In this way, the values stored in the database from one API can be transferred to serve as the arguments for another API.

\let\oldnl\nl
\newcommand{\nonl}{\renewcommand{\nl}{\let\nl\oldnl}}

\begin{algorithm}[!ht]
\renewcommand\baselinestretch{0.9}\selectfont
\caption{Mutation algorithm\Comment{\lingming{strange to have line numbers for inputs/defs; remember to change the line refs in the texts as well}}}
\label{alg:mutation}
\nonl\textbf{Input:}\\
\nonl\quad $API$ \ccc{the API under test to be mutated}\\
\nonl\quad $VS$ \ccc{\apivs}\\
\nonl\quad $DB$ \ccc{\argdb}\\
\nonl\textbf{Define:}\\
\nonl\quad $TypeRule$ \ccc{type mutation strategies}\\
\nonl\quad $ValueRule_{rand}$ \ccc{random value mutation strategies}\\
\nonl\quad $ValueRule_{db}$ \ccc{database value mutation strategies}\\
\SetKwFunction{FMain}{Mutate}
\SetKwProg{Fn}{Function}{:}{}
\Fn{\FMain{$API$, $VS$,\Comment{ $TS$\lingming{what is TS, it is never used!\david{type space, now switched to fuzztype}},} $DB$}}{
    \While {$notFinished$} {
        $entry = selectNext(VS[API])$ \label{lst:line:selectnext}\\
        $argNum = len(entry)$\ccc{number of arguments} \label{lst:line:argnum}\\
        $numMutation=Random.get(argNum)$ \label{lst:line:numMutation}\\
        \While{$numMutation > 0$} {
            $argIndex = selectNext(argNum)$ \label{lst:line:argindex}\\
            $argType = \fuzztypenocode(entry[argIndex])$ \label{lst:line:type}\\
            $argName = entry[argIndex].name$ \label{lst:line:argname}\\
            $argValue = entry[argIndex].value$ \label{lst:line:argvalue}\\
            \If{$doTypeMutation()$} {
                $argType=TypeRule(argType)$ \label{lst:line:typerule}\\
            }
            \If{$selectRandOverDB()$} {
                $next = ValueRule_{rand}(argType, argValue)$ \label{lst:line:randvalue}\\
            }
            \Else {
                $next = ValueRule_{db}(API, argType, argName, DB)$ \label{lst:line:dbvalue}\\
            }
            $entry[argIndex] = next$\\
            $numMutation = numMutation - 1$\\
        }
        $run(entry)$ \label{lst:line:run}\\
    }
}

\SetKwFunction{FMain}{$ValueRule_{db}$}
\SetKwProg{Fn}{Function}{:}{}
\Fn{\FMain{$API$, $argType$, $argName$, $DB$}}{
\Comment{    $APIs, similarities = db.query_{similar}(argType, API, argName)$\\
    $probs = softmax(similarities)$\\
    $API = pick\_by\_prob(APIs, probs)$\\
    $val = db.sample(argType, API, argName)$\\
    \algorithmicreturn{$val$}}
  $APIs=DB.query(argType, API, argName)$ \label{lst:line:apiquery}\\
  $\lbracket \apii, \simVar\rbracket=\simFunc(APIs, API)$ \label{lst:line:simapi}\\
    $\lbracket \apii, \probVar\rbracket=Softmax(\lbracket \apii, \simVar\rbracket)$ \label{lst:line:softmax}\\
    $API'=sample(\lbracket \apii, prob\rbracket)$ \label{lst:line:apiselected}\\
    $val = sample(DB, API', argType, argName)$ \label{lst:line:valueselected}\\
    \algorithmicreturn{ $val$}
    }
\end{algorithm}


\subsection{Test Oracle}
\label{subsec:testoracle}
In this phase, we leverage the following ways to resolve the test oracle problem and detect potential DL library bugs:
\Comment{\lingming{is it better to change diff testing into Wrong-Computation Bugs, and change meta testing into Performane Bugs as you want the three titles be at the same level?}
\david{Out of space. I move examples here into discussion part.}}

\parabf{Wrong-Computation Bugs.}
We consider three modes to run each API: CPU, GPU with CuDNN disabled, and GPU with CuDNN enabled.\Comment{ We observe that differential testing is a simple yet effective way to detect various kinds of bugs, especially wrong computation results.} In this way, we can detect wrong-computation results by comparing the results between different execution modes.

\parabf{Performance Bugs.}
\newcommand{\mach}{\mathcal{M}}
We leverage metamorphic relations~\cite{chen2018metamorphic,su2021fully} to detect performance bugs with \tech. \Comment{As discussed in Section~\ref{sec:background}, }More and more data types and hardware accelerators have been proposed in order to boost the DL library performance in recent years. Several floating point data types are specially designed for tensors, including \textit{float32}, \textit{float16}, \textit{tf32}, \textit{bfloat16}, which also appear in our aforementioned tensor data type system. We observe the fact that on the same machine (hardware) $\mach$, APIs with the same function arguments $args$ and tensors of the same shapes $tensor\lbracket n,DT\rbracket$\Comment{\lingming{use $\lbracket$ and $\rbracket$ for tuples globally to avoid confusion with lessthan/greaterthan symbols}\david{Yes.}} tend to hold the following metamorphic relationship in terms of time cost:
\begin{equation*}
\small
\begin{split}
precision(DT_1)&<precision(DT_2)\implies\\
cost(\mach, API, args, tensor\lbracket n,DT_1\rbracket)&<cost(\mach, API, args, tensor\lbracket n,DT_2\rbracket)
\Comment{if \quad&precision(DT_1) < precision(DT_2),\quad then\\
&time\_cost(m, API, args, tensor<n,DT_1>) \\
<\quad&time\_cost(m, API, args, tensor<n,DT_2>)}
\end{split}
\end{equation*}
This indicates that $DT_1$ carrying less precision information than $DT_2$ tends to\Comment{\lingming{changed the last `should' to `tend to'; should change this one as well as time diff is really very flaky}} execute faster. For instance, $DT_1$ can be \textit{float16} while $DT_2$ is \textit{float32}, as long as the API supports both data types of tensors.
\Comment{
Apart from that, more advanced hardware/backend should be faster during execution, which can be formalized as follows:
\begin{equation}
\begin{split}
if \quad&adv(m_1) > adv(m_2),\quad then\\
&time\_cost(m_1, API, args, tensor<n,DT>) \\
<\quad&time\_cost(m_2, API, args, tensor<n,DT>)
\end{split}
\end{equation}
where machine $m_1$ more advanced than machine $m_2$.
}

\parabf{Crash Bugs.}
If an API crashes or throws runtime exception, then it may be considered as a crash bug. 
Meanwhile, it could also be due to invalid test inputs which can be generated during the fuzzing process. To automatically filter out such false alarms, we build scripts to heuristically remove crash bugs which throw meaningful exceptions on all backends for invalid inputs, e.g., `\CodeIn{ValueError}', `\CodeIn{InvalidArgumentError}', etc.
\Comment{Manual inspection is needed to further judge whether such crashes are due to invalid test input or not (according to the API documentation), because our mutation strategies do not guarantee that all the mutants are valid.
There are around 80\% crash bugs which are actually due to invalid input. The reason is that our mutation algorithm does not guarantee that the input is valid. We have a script to heuristically remove crash bugs which throw meaningful exceptions for invalid input, e.g., `ValueError', `InvalidArgumentError', etc.}
\Comment{
\lingming{keep in mind that we need to show the false positive rate of \tech in the last rq}\david{How about we remove text like 'manual inspection' here? Also, confirming crash bugs can also use `differential testing' based on failure message. For example, if it crashes only on GPU, then it is indeed a crash bug. If it crashes both on GPU and CPU with the same failure message, it is likely due to invalid input. I have put the explaining text there in the beginning of RQ5. I do not want to use word like `false positive'. Let's say `invalid input' instead.}. }As a result, if the test program crashes (e.g., segmentation fault),\Comment{ throws unexpected exception,} or throws unexpected exception for valid inputs on certain backend(s)\Comment{ (\david{according to documentation})}, it is considered as a crash bug.

\Comment{\textbf{Missing Checking}\qquad
If an invalid test input is generated for an API, but the API runs silently and successfully without throwing any exception or warning, it is considered as a \textit{Missing Checking} bug. Such bugs are harmful because APIs are wrongly invoked without warning users of the bug. This is a new category of bugs unreported by prior work \cite{lemon,cradle}. Like crash bugs, missing checking bugs require manual inspection to see whether the test input is valid or not according to documentation.
}












\section{Experimental Setup}
In the study, we address the following research questions:
\begin{itemize}
\item \textbf{RQ1:}\Comment{What is the coverage breakdown of three different sources of input?} How do the three different input sources of \tech{} (without any mutation) contribute to DL library testing?
\item \textbf{RQ2:}\Comment{ Do mutants generated by \tech{} effectively invoke more code in DL libraries?} How does \tech{} with different numbers of mutations for each API perform for DL library testing?
\item \textbf{RQ3:}\Comment{ How do type mutations and value mutations contribute to coverage during fuzzing?} How do different mutation strategies impact \tech{}'s performance?
\item \textbf{RQ4:} How does \tech{} compare with existing work?
\item \textbf{RQ5:} Can \tech{} detect real-world bugs?
\end{itemize}

Our experiments are mainly performed on the stable release versions of\Comment{popular} DL libraries: \pt 1.8 and \tf 2.4. The machine for running experiments is equipped with Intel Xeon CPU (4 cores, 2.20GHz), NVIDIA A100 GPUs, Ubuntu 16.04, and Python 3.9.

\subsection{Implementation}
\label{subsec:implementation}
\Comment{\lingming{merge this into the implementation subsec as the first phase. also, add the implementation details for the 3rd fuzzing phase, e.g., what DB you used and how you implemented the algorithms}}
\parabf{Code/Model Collection.} Code/model collection is essential to form the original seed test pool for our fuzzing technique. To build an extensive pool, for documentations, we download all \numdocpt/\numdoctf pieces of code snippets from the official documentations of \pt/\tf. More specifically, we use the \CodeIn{bs4} Python package~\cite{bs4} to automatically parse the documentations to obtain the code snippets.
\Comment{\lingming{how many did we download in total and how many we can use here? Can you also talk a bit more about how you implemented the crawler (e.g., using what language, what libs/frameworks, and how to parse the docs to get the code snippets)? }
\david{
This is done in a automated way by parsing the HTML webpages using a Python package \CodeIn{bs4}~\cite{bs4}.
}}
Note that not all code snippets crawled from documentations are immediately executable. Thus we also build a simplistic repair tool to insert omitted code in the examples (e.g., \CodeIn{import} sections) to make more code snippets executable.
For developer tests, we run all existing Python tests for \pt by running \CodeIn{python run\_test.py} in the test directory, while for \tf we run all python files with suffix \CodeIn{\_test.py}.\Comment{ \david{For both \pt and \tf, tests are written in files with suffix \CodeIn{\_test.py}, and we run execute tests by running all the files.}. \lingming{what unit test frameworks and/or build systems do \tf and \pt used to run the tests?}. \yinlin{In theory, we can use bazel to run tensorflow developer tests, however, we need to do instrumentation in python code, but I am not sure if we can do so with bazel test.}}
\Comment{Some tests in \pt and \tf are written in Python by the developers, while some are written in C++. Because \tech{} only instruments and tests the APIs in Python, thus we only run Python tests \textit{without} runing C++ tests.}
For DL models, we obtain a diverse set of \numModelsPtAll/\numModelsTfAll DL models from official model zoos of \pt/\tf, and popular GitHub repositories. The detailed information about the models can be found in our repository~\cite{freefuzzrepo}. \Comment{Remove TABLE! Table~\ref{tab: modelcompare} shows the detailed categorization of models used for various tasks, compared against models used in the prior \lemon and \cradle work. We can observe that our dataset includes a much more diverse model set than the prior work.} \Comment{Substantial efforts are devoted into collecting a diverse set of models. Setting up proper environment and running them successfully is by no means easy. Some models require us to download huge dataset manually. Moreoever, some code from the wild is not fully compatible with the newest version of the dependencies that we installed using pip, which requires additional efforts to fix their code manually. It can take up minutes to even several hours to make one model execute successfully.}

\Comment{
\begin{table}\centering
\caption{Categorization and Comparison of Models as Input\Comment{\lingming{follow this style and use scalebox for all other result tables}}\lingming{maybe remove this table due to space limit}}\label{tab: modelcompare}
    \scalebox{0.86}{
\begin{tabular}{|l|c|c|c|c|}
\hline
\multirow{2}{*}{} &\multicolumn{2}{c|}{\tech} &\multirow{2}{*}{\lemon} &\multirow{2}{*}{\cradle} \\
\cline{2-3}
&PyTorch &TF & & \\
\hline\hline
Image Classification & 30 & 24 & 10 & 26 \\
Superresolution & 5 & 3 & - & - \\
Optical Character Recognition & 2 & 1 & - & \\
Style Transfer & 3 & 2 & - & \\
Time Sequence Prediction & 5 & 4 & 2 & 3 \\
Text Classification & 11 & 5 & - & - \\
Language Modeling & & 6 & - & - \\
Machine Translation & 4 & 5 & - & - \\
Question and Answering & 1 & 7 & - & \\
Text Generation & & 8 & - & - \\
Reinforcement Learning & 10 & 11 & - & 1 \\
Generative Models & 10 & 11 & - & - \\
Graph Neural Network & 2 & 6 & - & \\
Adversarial Attacks & 11 & 1 & - & \\
Bayesian Learning & 7 & 2 & - & \\
Other Applications & 1 & 4 & - &
\\
\hline
\# Models & 102 & 100 & 12 & 30 \\
\hline
\end{tabular}
}
\end{table}
}

\Comment{\parabf{Code Collection.}  We observe that not all code from documentation are immediately executable, so we need to add the ommitted \CodeIn{import}. Take the code shown in Figure~\ref{fig:conv-code} for instance, we need to add \CodeIn{import torch} and \CodeIn{import torch.nn as nn} so that it can work. We first manually fix a few examples, then we find some common patterns so that adding \CodeIn{import} can be largely done in an automated way afterwards.}

\parabf{Instrumentation.}\Comment{ There are two types of APIs for popular DL libraries, one is called class APIs (e.g., \CodeIn{torch.nn.Conv2d} shown in Figure~\ref{fig:conv-code}), and the other is function APIs (e.g., \CodeIn{torch.nn.functional.conv2d}). The major difference between the two is that invoking a class API creates an object and the object expects tensors as arguments, while invoking a function API directly expects tensors as part of its function arguments. Some slight differences exist in instrumenting these two types of APIs.} 
We get the lists of all Python APIs from official documentation of \pt and \tf, and hook them in \CodeIn{\_\_init\_\_.py}  (a file for a package that will be automatically executed if the package is imported) in the root of the library package by adding a wrapper for each API in the list\Comment{function or class in the API list}. This is done transparently and fully automatically for the users so that they do not need to modify any of their code (model code) for instrumentation. In this way, \numHookAPIsPt APIs from \pt and \numHookAPIsTftwo APIs from \tf are instrumented for dynamic value tracing. Furthermore, we leverage the MongoDB database~\cite{mongodbwebsite,pymongodoc} to record \apivs and \argdb.

\parabf{Mutation.}\Comment{\david{We implement Algorithm~\ref{alg:mutation} for mutation with standard Python packages. The input arguments (\apivs and \argdb) are simply stored in comma-separated values (CSV) files. We set the random function \CodeIn{doTypeMutation()} to return \CodeIn{True} with a probability of 1/3 and \CodeIn{selectRandOverDB()} to return \CodeIn{True} with a probability of 2/3. We find that the overhead largely stems from running the mutant on hardware, and the running cost of other parts of the algorithm is almost negligible.}} We implement our main Algorithm~\ref{alg:mutation} for mutation with standard Python packages. The implementation details can also be found in our project repository~\cite{freefuzzrepo}. \Comment{The input arguments (\apivs and \argdb) are simply stored in comma-separated values (CSV) files. We set the random function \CodeIn{doTypeMutation()} to return \CodeIn{True} with a probability of 1/3 and \CodeIn{selectRandOverDB()} to return \CodeIn{True} with a probability of 2/3.\lingming{reviewers will ask why 1/3 why not 1/2, 1/4, 1/5? better just use random choice, i.e., 1/2} We find that the overhead largely stems from running the generated mutants on hardware, and the running cost of other parts of the algorithm is almost negligible.}

\parabf{Test Oracle.}
The implementation of differential testing is simple. The example code for \pt is shown in Figure~\ref{fig:difftestconv}.\Comment{ As the assertion is for tensors, we use \CodeIn{torch.allclose} (resp. \CodeIn{tf.debugging.\allowbreak assert\_equal}) for tensor comparison for \pt (resp. \tf).} Meanwhile, the implementation of metamorphic testing is to wrap the invocation of APIs with code for timing.

\subsection{Metrics}
To thoroughly evaluate \tech, we use the following metrics:

\parabf{Number of Covered APIs.}
\Comment{To measure the number of covered Python APIs, \tech first hooks a list of \numHookAPIsPt (resp. \numHookAPIsTftwo) from \pt (resp. \tf) from the official websites.}
Due to the large number of APIs in DL libraries, it is of great importance to show the number of covered APIs as an important metric of testing adequacy. Surprisingly, such an important metric has been largely overlooked by prior work on DL library testing~\cite{lemon, cradle}.
\Comment{After executing some code/models, if certain APIs are covered, then some entries will exist in the value space of the covered APIs. APIs with non-empty value space are considered as covered.}
\Comment{\lingming{this is not quite right. you just repeat something we already presented elsewhere. for here, it is more important to explain why the number of covered apis is important for measuring fuzzing effectiveness.}

\david{We may want to abandon \#VS for comparion with LEMON/CRADLE. This index is hard to make it meaningful, especially now we can get coverage.}\lingming{agree. you can still list value space (no need to list value space/apis as that basically the same as value space), but at the same time claim this just shows some numbers about the traced items and cannot indicate fuzzing effectiveness.}}

\parabf{Size of Value Space.} Each API invocation can add one entry into the \apivs. Therefore, we use the total size of \vs for all APIs to serve as the metric to analyze and compare different input sources. To be more accurate, we count the number of entries in the \apivs after removing duplicate entries. Please note that this is just used to show the scale of the traced data, and does not necessarily indicate fuzzing effectiveness.

\Comment{
For each API, we differentiate invocations by parameters or input tensors' shapes and data types. For example, although running the code from Figure~\ref{fig:conv-code} twice will produce different computation results (due the the randomly initialized values of the input tensor), yet the values of API argument and the shapes or data types of tensors do not differ, and thus the two invocations will be counted as only once in terms of \textit{distinctive} API invocation. Based on how \tech records the \textit{value space} of each API, we are able to remove the redundant entries in the value space of each API. Adding together the number of entries in the value space of each API is the total size of value space of all covered APIs. This metric indicates how diverse the input to APIs is.
}

\parabf{Line Coverage.} Code coverage is a widely adopted test adequacy criterion for testing traditional software systems~\cite{brosgol2011178c, zhang2018hybrid, legunsen2016extensive} and even the recent tensor compilers~\cite{liu2022coverageguided}. For example, it is impossible for a test to detect bugs in code portions without executing them. Surprisingly, although state-of-the-art DL library testing techniques (e.g., \lemon) claimed to invoke more library code~\cite{lemon}, they did not report any code coverage in their experiments.\Comment{For more rigorous/thorough evaluation of DL library testing, } We spent tremendous time and efforts setting up the environment for collecting the most widely used line coverage via GCOV~\cite{gcov} for both \pt and \tf. More specifically, we even fixed a bug in the Bazel build system~\cite{bazel} used for building \tf to perform coverage collection.\Comment{\lingming{so we only traced C/C++ cov and no Python cov right? do we want to add a reason for why only tracing C/C++ cov here?}
\david{We only traced C++ cov. No strong reason. Actually Python coverage is even higher than C++...}\lingming{where do developers typically fix the bugs we report? in Python or C code? if mostly C, then we have some good reason.  }\david{typically fix in Python.}
}
Note that we only trace C/C++ code coverage because the C/C++ implementation provides the fundamental support for operators in DL libraries and almost all the high-level Python APIs finally invoke the C/C++ code.


\parabf{Number of Detected Bugs.} Following prior work on software testing in general and DL library testing~\cite{chen2020practical, lemon, cradle}, we also report the number of actual bugs detected for the studied DL libraries. \Comment{Due to the fact that code coverage may not be directly The number of detected bugs is good metric to measure the capability of bug-finding.}
\section{Result Analysis}
\label{sec:resultanalysis}

\subsection{RQ1: Input Source Study}
\label{subsec:inputsourcestudy}
In this RQ, we aim to study the effectiveness of directly applying \tech's traced dynamic information (without any mutation) for testing DL libraries. The main experimental results are shown in Table~\ref{tab:coveragebreakdown}, where we explore different settings, including using documentations only, tests only, models only, and all information together for both \tf and \pt. For each setting, we show the number of covered APIs (Row ``\# API''), the number of traced unique API invocations (Row ``\# VS''), and the line coverage achieved when directly running the traced API invocations (Row ``Line Cov.''). From the table, we can observe that different sources of information all tend to be helpful for testing DL libraries. For example, although the test information covers the least number of APIs for \tf, it can still help directly cover \numtestapitf APIs and 31293 lines of code; similarly, although the model information covers the least number of APIs for \pt, it can still help directly cover 145 APIs and 26292 lines of code. Also, another interesting observation is that the settings covering more APIs tend to also achieve higher code coverage.\Comment{The reason is different APIs have distinctive functionalities, and invoking them should cover different code.} The reason is that different APIs usually implement different functionalities, and thus usually cover different DL library behaviors/paths. This actually also demonstrates the effectiveness and necessity of API-level testing for DL libraries since it is much easier to cover more APIs at this level than traditional model-level DL library testing~\cite{lemon, cradle}. 

\begin{figure}
    \centering
    \includegraphics[keepaspectratio=true,width=\columnwidth]{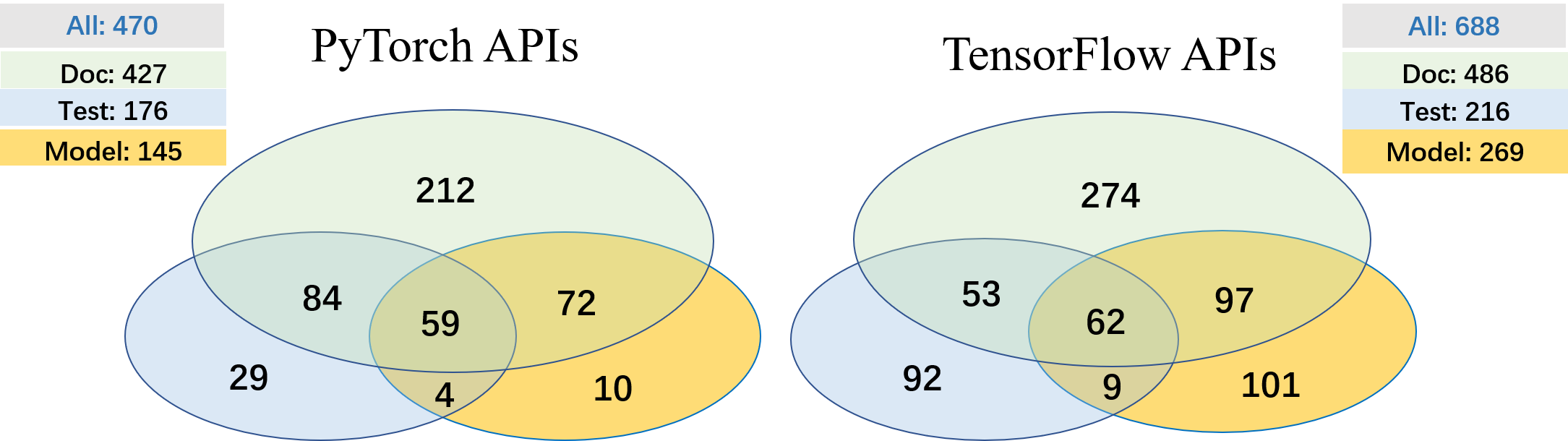}
    \precaptionspace
    \caption{Venn diagram for covered APIs\Comment{better to have separate figs for API and cov? putting them together is a bit hard to read}\label{fig:vennAPI}}
    
    \centering
    \includegraphics[keepaspectratio=true,width=\columnwidth]{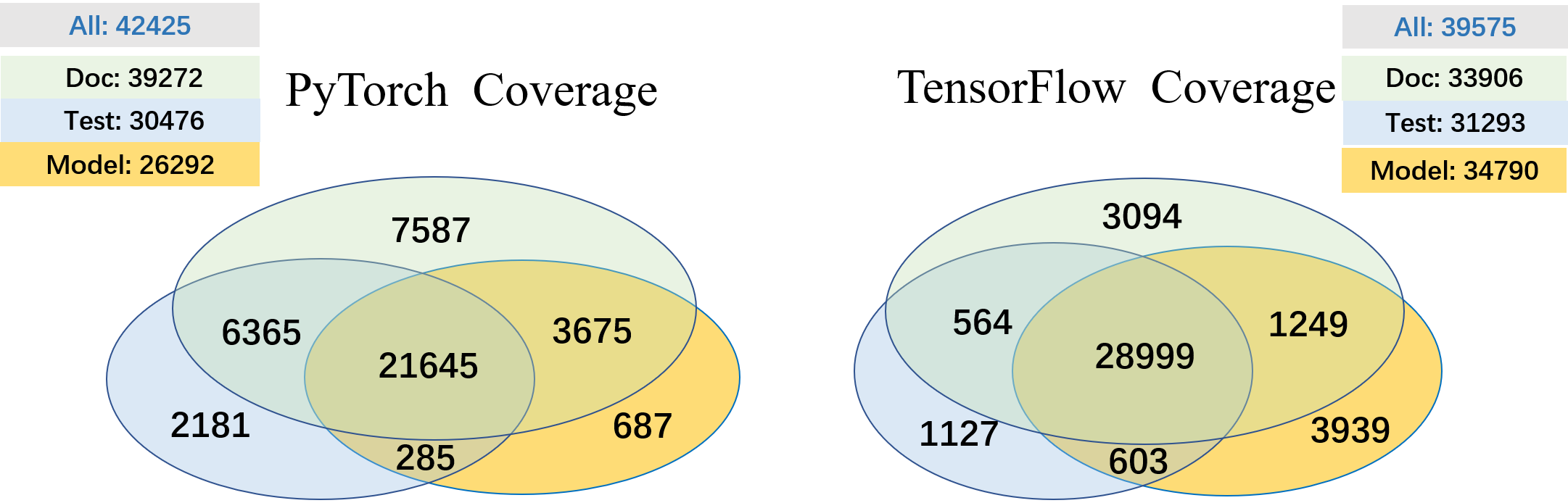}
    \precaptionspace
    \caption{Venn diagram for code coverage\label{fig:venncoverage}}
    
\end{figure}

    We can also observe that putting all three sources of information together can achieve even better results than using any single source of information. For example, it can cover \numTracedAPIsPt/\numTracedAPIsTf APIs for \pt/\tf, and 42425/39575 lines of code for \pt/\tf. To better analyze the contribution of each source of information, we further leverage the Venn diagrams in Figure~\ref{fig:vennAPI} and Figure~\ref{fig:venncoverage} to present the detailed breakdown of the number of covered APIs and coverage respectively. From the figure, for both \tf and \pt, each source of inputs exclusively covers some APIs and only a small number of APIs are covered by all three sources of information. For example, only 59/62 out of all the \numTracedAPIsPt/\numTracedAPIsTf covered APIs are covered by all three sources of inputs on \pt/\tf. Meanwhile, although each source of inputs still exclusively covers different code portions, the majority of covered code tends to be shared by all three sources of inputs. The reason is that although different APIs implement different code logic, they can be decomposed to a set of common low-level operators implemented in C/C++. Overall, the experimental results further confirm that it is necessary and important to consider different sources of information for effective DL library testing. 

\Comment{Note that }Tracing the three sources of inputs is a \emph{one-time effort} and can be used for testing all subsequent versions of the same DL libraries. Meanwhile, it is also important to demonstrate that the time overhead is acceptable and not extremely high. Therefore, we further discuss the overhead for constructing the three sources of inputs. For the documentation source, the code snippets are usually quite short and fast to run. In total, \tech{} takes less than 20 min for tracing all the documentation code snippets for both \tf and \pt. 
\Comment{For example, running \numdocpt pieces of code from the \pt documentation covers \numdocapipt APIs, producing \numdocapivspt different entries in the value space of APIs.
The reason why the number of code snippets does not match the number of covered APIs is because there exist overlapping APIs covered in different code snippets.}
For the developer tests, tracing the \numtestpt/\numtesttf official tests written by developers from \pt/\tf consumes about 2.5/5.0 hours.
Lastly, for the model source, \tech runs\Comment{ \chenyuan{runs}} all the \numModelsPtAll/\numModelsTfAll models stated in Section~\ref{subsec:implementation} with instrumentation for \pt/\tf, consuming less than 1 hour for each of them.\Comment{ \lingming{lingming will restructure the rq again in the next pass (after all comments/questions get answered)}}\Comment{More specifically, we run inference for all models, like prior work~\cite{lemon,cradle}. To save time, we only need to ensure that the APIs defined in the model are invoked at least once, and there is no need to run the inference for the whole dataset because iterating over the dataset simply repeats the invocations of APIs. Therefore, running one model takes no more several seconds.}\Comment{\lingming{i assume the time for test and model execution should include instrumentation? if so, make it clear}}

\Comment{Running code snippets collected documentation (sequentially) has low overhead, consuming around 20 minutes for \pt and around 19 minutes for \tf.
For \pt, running \numdocpt pieces of code from documentation covers \numdocapipt APIs, with a total of \numdocapivspt different entries in the value space of APIs, (summing up all the distinctive entries in the value space of each API).
The reason why the number of code snippets does not equal to the number of covered APIs is because there exist some overlapping APIs covered in different code snippets.
Besides code from official documentation, \tech also run official tests written by the developers of DL libraries. Running the \numtestpt (resp. \numtesttf) official tests written by developers from \pt (resp.\tf) consumes about 2.5 (resp. 5) hours.
Lastly, \tech run \numModelsAll models stated in Section~\ref{subsec:implementation}. More specifically, we run inference for all models, like prior work~\cite{lemon,cradle}. To save time, we only need to ensure that the APIs defined in the model are invoked at least once, and there is no need to run the inference for the whole dataset because iterating over the dataset simply repeats the invocations of APIs. Therefore, running one model takes no more than a minute.
The number of covered APIs, the total size of \apivs, and the number of covered lines in shown in Table~\ref{tab:coveragebreakdown}. The detailed breakdown of the number of covered APIs are shown in the venn graph in Figure~\ref{fig:vennAPI}. We can see that only a small number of APIs are in the intersection of all the three sources of input, and they cover the a large number of lines. Still, each source of input has their own advantages in terms of both the number of covered APIs and the line coverage.}

\Comment{
\tech forms a database  which stores the value space of every covered API for \pt and \tf respectively.
Table~\ref{tab:coveragebreakdown} shows the number of covered APIs (\texttt{\# API}), the number of entries in the value space for APIs (\texttt{\# VS}), and their ratio (\texttt{\#VS / \#API}). The columns (\texttt{Doc}, \texttt{Test}, \texttt{Model}) corresponds to the aforementioned sources, and the numbers in column \texttt{All} indicate the data in the database (combination of the three sources). Note that the numbers are collected after redundancy removal.
}

\begin{table}\centering
\caption{Statistics about different sources}\label{tab:coveragebreakdown}
\scalebox{0.8}{
\begin{tabular}{|l|cccc|cccc|}
\hline
\multirow{2}{*}{} &\multicolumn{4}{c|}{\tech PyTorch} &\multicolumn{4}{|c|}{\tech TensorFlow} \\\cline{2-9}
&Doc &Test &Model &All &Doc &Test &Model &All \\\hline
\hline
\# API &\numdocapipt &\numtestapipt &\nummodelapipt &\Comment{476}\numTracedAPIsPt &486 &\numtestapitf &\nummodelapitf &\Comment{701}\numTracedAPIsTf \\
\# VS &\numdocapivspt &\numtestapivspt &10898 &15532 &1810 &6879 & 36638 & 45269 \\
Line Cov. & 39272 & 30476 & 26292 & 42425     & 33906 & 31293 & 34790 & 39575\\
\hline
\end{tabular}
}
\end{table}


\Comment{
\subsection{RQ2: Comparison on coverage of input with \lemon and \cradle}
}

\subsection{RQ2: Coverage Trend Analysis}
\begin{figure}
    \centering
    \includegraphics[keepaspectratio=true,width=0.8\columnwidth]{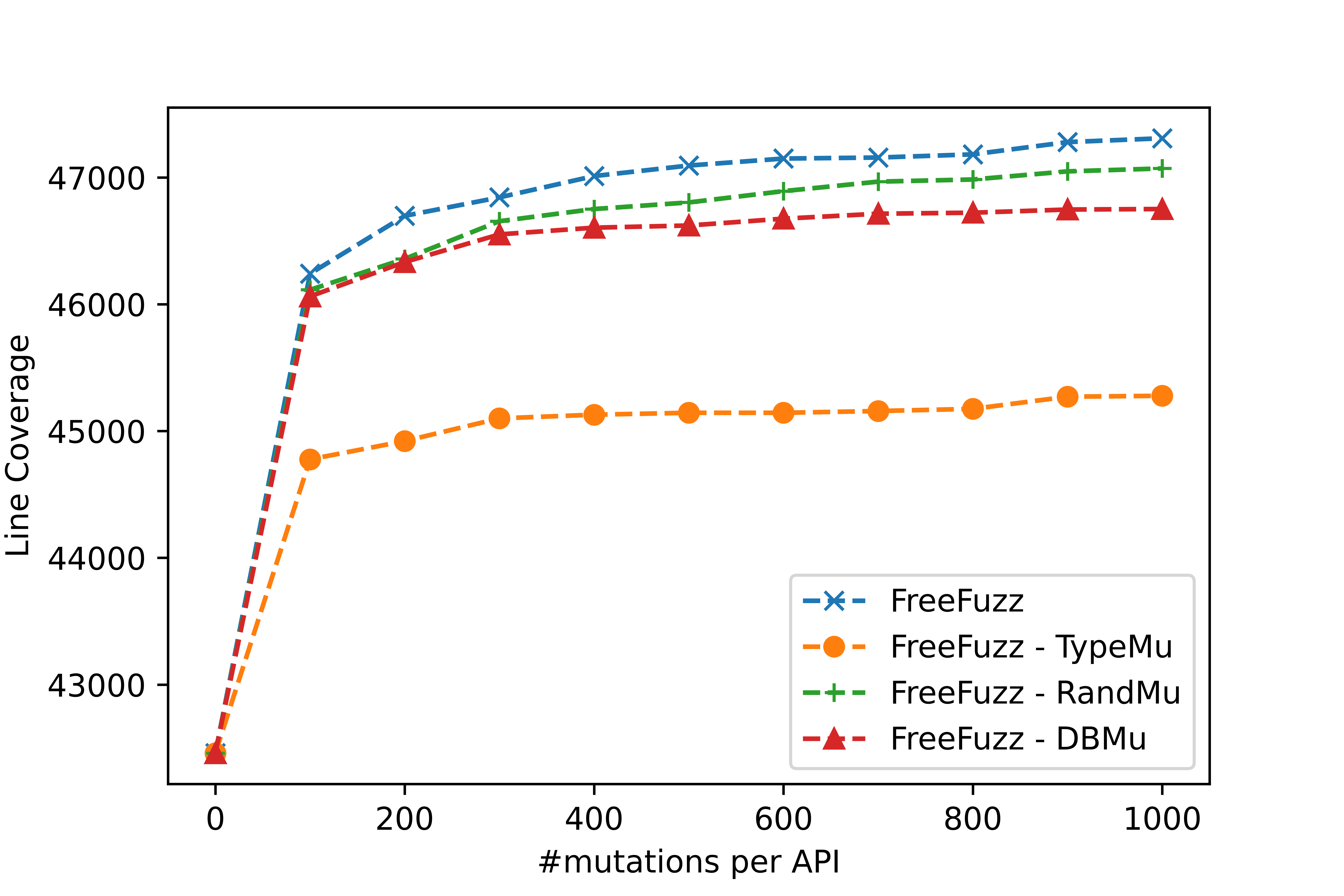}
    \caption{Coverage trend analysis for \pt \label{fig:trendpytorch}}
    \centering
    \includegraphics[keepaspectratio=true,width=0.8\columnwidth]{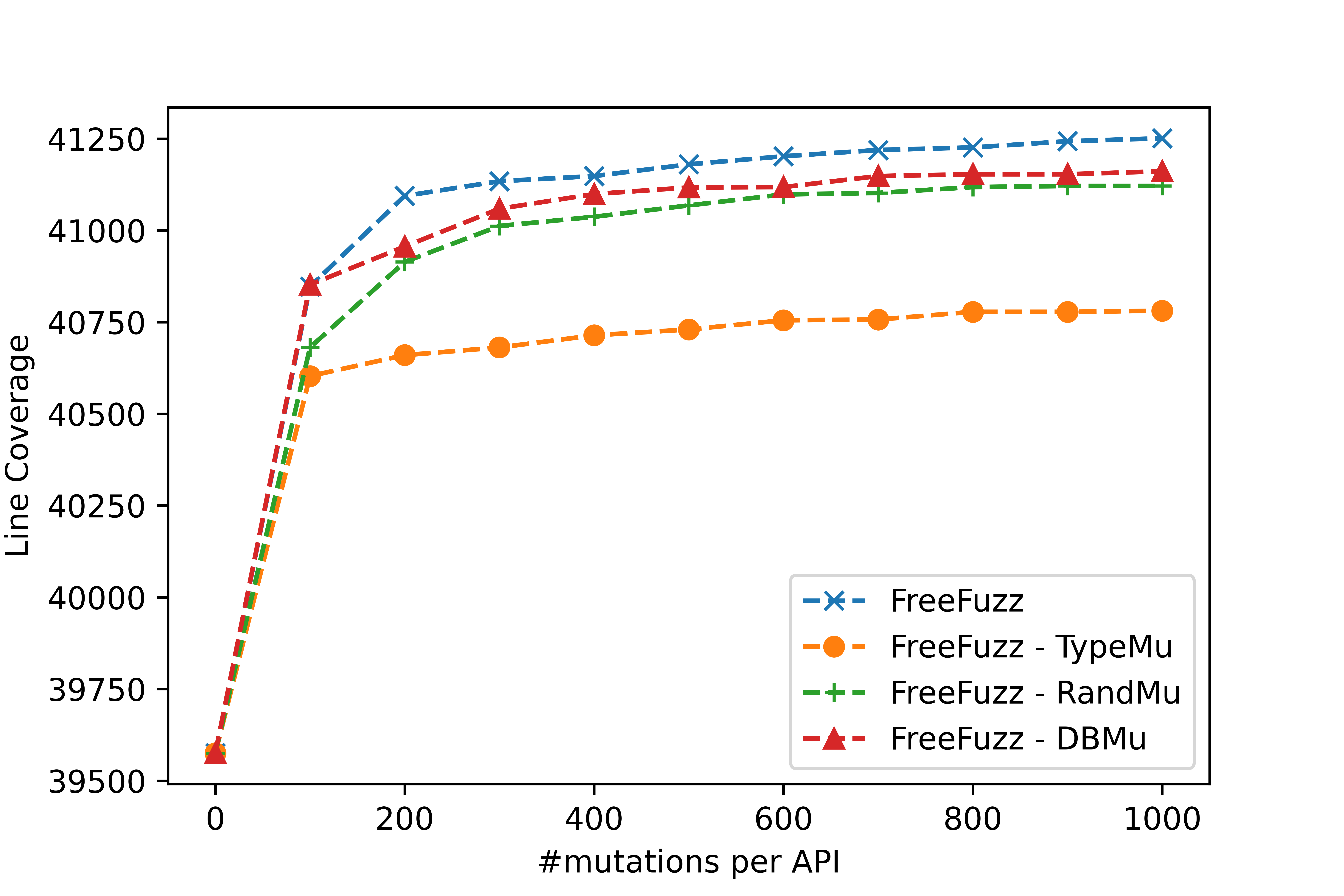}
    \caption{Coverage trend analysis for \tf\Comment{\lingming{the figs need to be redrawn with same style. The fonts for \pt are too small while \tf is blur}\david{improved}}\label{fig:trendtensorflow}}
    
\end{figure}

In this RQ, we present the effectiveness of \tech with different numbers of mutations for each API under test. The experimental results are shown as the blue lines (with legend ``\emph{\tech}'') in Figure~\ref{fig:trendpytorch} and Figure~\ref{fig:trendtensorflow}, where the \emph{x} axis presents the number of mutants generated for each API (from 100 to 1000 with the interval of 100) while the \emph{y} axis shows the overall coverage achieved via generating different number of mutants for each API (the union of all coverage sets for all tested APIs). Note that the start point denotes the code coverage achieved by directly executing the original test inputs traced without any mutation. From the figure, we can observe that for both \pt and \tf, \tech can indeed cover more lines of code with more mutations enabled for each API under test, demonstrating the overall effectiveness of our mutation strategies. Furthermore, we can also observe that the coverage becomes largely stable after running 600 mutations for each API\Comment{for both \tf and \pt}, indicating that 600 mutations can be a cost-effective choice in practice.
Regarding the time cost, the total running time for generating and running all 1000 mutants for all APIs is 7.3 hours for \pt and 9.9 hours\Comment{\lingming{i remember chenyuan mentioned that we are now much faster, shall we update this time info?}} for \tf\Comment{, demonstrating the efficiency of the proposed approach}. Note that such overhead is totally acceptable for fuzzing, e.g., traditional binary fuzzing techniques are usually applied for 24h~\cite{bohme2017coverage} and\Comment{state-of-the-art DL library fuzzing work} \lemon takes over 24h~\cite{lemon}. 
\Comment{We measure the trend of code coverage and it is shown in Figure~\ref{fig:trendpytorch} (the curve with crossing annotated with `full'). We mutate each API 1000 times (with our mutation algorithm). In order to show the trend of coverage, we get the code coverage for the whole DL system every time we mutate all APIs 10 times. Thus, if different APIs cover the same code, the same covered code is only counted once because the coverage is measured for the whole system.
The total running time is \Num{X} minutes for \pt and \Num{Y} minutes for \tf.
From the figure, we can find that our fuzzing approach can indeed cover more lines of code, and the coverage of the DL library is largely stable after running 600 mutations for each API.}

\newcommand{\noType}{\tech-TypeMu\xspace}
\newcommand{\noRand}{\tech-RandMu\xspace}
\newcommand{\noDB}{\tech-DBMu\xspace}
\newcommand{\noAny}{\tech-AllMu\xspace}

\subsection{RQ3: Different Mutation Strategies}
\label{subsec:variant}
After tracing the initial inputs from various sources, \tech performs three different mutation strategies in tandem (as detailed in Algorithm~\ref{alg:mutation}). Therefore, in this RQ, we further study the impact of each mutation strategy by disabling it. To this end, we have three \tech variants, \noType (disabling the type mutation strategy), \noRand (disabling the random value mutation strategy), \noDB (disabling the database value mutation strategy). Note that we also include the variant that does not perform any mutation at all, i.e., \noAny. The experimental results for all the studied variants with different number of mutations for each API are also shown in Figure~\ref{fig:trendpytorch} and Figure~\ref{fig:trendtensorflow}. Note that the start point for all other variants denotes the coverage achieved by \noAny. From the figure, we can have the following observations. First, the default \tech outperforms all the other studied variants, indicating the importance and necessity of all the three mutation strategies of \tech.\Comment{ For example, the default \tech can achieve \lingming{?/? higher coverage}\david{yes} compared with the second best variant for \pt/\tf.} Second, we can also observe that random-value and database-value mutation strategies perform similarly in terms of code coverage, while type mutation can be even more effective since the low-level implementations for different types tend to be more different. \Comment{The reason is that the low-level implementations for different parameter types tend to be different, and the type mutation strategy of \tensordtype can effectively explore them\lingming{check it}.} \Comment{\lingming{explain why type mutation can trigger more cov}\david{the low-level implementations for tensors with different data types are different, and the type mutation strategy of \tensordtype can effectively explore them.}}

\Comment{We measure the effectiveness of our mutation strategies by comparing with 3 variants of the full version of Algorithm~\ref{alg:mutation} and another variant without muation: Variant 1: Full - TypeMutation (i.e., removing the type mutation part from the full approach of \tech{}, shortened as `full - type'); Variant 2: Full - RandomValue (full - random); Variant 3: Full - DatabaseValue (full - database); Variant 4: No Mutation. We measure the coverage of the whole system for Variant 1, 2, 3 by generating 1000 mutants for each API. For Variant 4, we run all the recorded invocations in the \apivs. The trend is shown in Figure~\ref{fig:trendpytorch}. Note that variant 4 (no mutation) is the starting point of all the other three variants. We find that variant 1 has the lowest coverage, meaning that type mutation is the most effective in achieving more coverage. Also, each variant is strictly lower than the full approach, so all the mutation strategies do contribute to fuzzing in terms of coverage.}

\Comment{
\begin{table}[!htp]\centering
\caption{Coverage analysis of different mutation variants}\label{tab:variant}
\begin{tabular}{c|c|c}
& \pt &\tf \\
\hline
Variant 1 & &  \\
Variant 2 & &  \\
Variant 3 & & \\
Variant 4 & & \\
\hline
\end{tabular}
\end{table}
}

\begin{table}\centering
\caption{Comparison on input coverage}\label{tab:coverinput}
\scalebox{0.85}{
\begin{tabular}{|l|c|c|c|}
\hline
&\tech (tf1.14 full) &\lemon &\cradle \\
\hline\hline
\# API & 313 & 30 & \numcradleAPI \\
\# VS & 9338 & 1808 & 2893 \\
Line Cov. & 33389 & 29489 & 28967 \\
\hline
\end{tabular}}
\end{table}


\begin{table}\centering
\caption{Comparison with LEMON on mutation}
\label{tab:mutatecomparisonlemon}
\scalebox{0.85}{
\begin{tabular}{|l|c|c|c|}
\hline
& \tech (tf1.14 full) &\tech (models only) &\lemon  \\
\hline\hline
\# API & 313 & 30 & \numlemonAPI \\
\# VS & 305463 & 913 & 7916\\
Line Cov. & \linecovFreeComp & 30488 & \linecovlemon \\
Time & 7h & 20 min & 25h\\
\hline
\end{tabular}}
\end{table}

\subsection{RQ4: Comparison with Prior Work}
In this RQ, we aim to compare \tech with the state-of-the-art \lemon~\cite{lemon} and \cradle~\cite{cradle} work for DL library testing.\Comment{\lingming{we need to explain somewhere in the paper why we only compare against lemon and cradle but not other ones like predoo (forgot its name)}. \david{\predoo only targets \predooOperatorNum APIs from \tf and focuses on precision testing, which is not a comprehensive tool to detect bugs for DL library, so we do not compare with it.}}
We first compare their sources of inputs in terms of the number of covered APIs and coverage.\Comment{ As stated in Section~\ref{subsec:implementation},} \lemon only uses \lemonModelNum models, \cradle uses \cradleModelNum models, and \tech{} considers three different sources of input with many more models in the wild. Since both \lemon and \cradle use \keras without supporting \pt, the comparison here is only conducted on \tf. Also, due to the fact that \lemon and \cradle do not support \tf 2 (used in our earlier experiments), we apply \tech on an old \tf version v1.14. For a fair comparison with prior work, we enforce \tech{} to use exactly the same models from \lemon as the DL model input source. To prepare the other two input sources for \tech, we collect developer tests and documentation code for \tf v1.14. The experimental results are presented in Table~\ref{tab:coverinput}: Column ``\tech (tf1.14 full)'' \Comment{\chenyuan{This should be 1.14}} simply runs the inputs traced by running the same models from \lemon as well as documentation code and tests from \tf v1.14; Columns ``\lemon'' and ``\cradle'' simply run the input DL models used in their original work. From the table, we can observe that, when no mutations are allowed, the input sources used by \tech can achieve much higher API and code coverage than \lemon and \cradle. 

We next study the effectiveness of the mutation strategies used by \tech and existing work (i.e., \lemon because \cradle performs no mutation). We follow the same methodology as the original \lemon work~\cite{lemon} when running its model-level mutations. For \tech{}, we also use the default setting, i.e., generating and running 1000 mutants for each covered API. The experimental results are shown in Table~\ref{tab:mutatecomparisonlemon}. Note that besides the default \tech and \lemon, Column ``\tech (models only)'' further includes the results of \tech with only the models from \lemon (without documentation code and developer tests) as the input for a more thorough comparison with \lemon. From the Table, we can observe that the default \tech can cover $\sim$\APImorecovered more APIs than \lemon while consuming $\sim$\overheadlowerfull less time! Although the coverage improvement is not as significant as the number of covered APIs, \tech can still outperform \lemon by $\sim$\morecoveragelemon. Also, surprisingly, \lemon only covers 5 more APIs via various model mutations compared to the original models, since only 5 unused layers preserve the strict input-output shape constraints imposed by \lemon and are added into the mutated models. Furthermore, \tech{} with models only can already outperform \lemon in terms of code coverage\Comment{\lingming{explain why we only have 30 apis while lemon has 35?}\david{The reason why \lemon covers 5 more APIs during mutation is because 5 unused layers (which preserve the input-output shape of tensors) are added into the mutated models according to the \lemon's mutation rules designed by experts, thus invoking 5 more APIs in model-level mutation, while \tech does not invoke APIs uncovered in sources of input.}} within 20min, i.e., \overheadlowerlemonmodel faster than \lemon! This further demonstrates the benefits of API-level testing compared with model-level testing\Comment{ for DL libraries}.
\Comment{various numbers in this paragraph looks quite impressive, and should go to abs and intro!}

\Comment{We first compare the sources of input in terms of the number of covered APIs and coverage. As stated in Section~\ref{subsec:implementation}, \lemon only uses \lemonModelNum models, \cradle uses \cradleModelNum models, and \tech{} considers three different sources of input with various models in the wild. Because both \lemon and \cradle use \keras without supporting \pt, so the comparison is only conducted on \tf. Due to the fact that \lemon and \cradle does not support \tf 2, we apply our mutation techniques on an old \tf version v1.14. \lemon does not consider documentation code or library tests, therefore, we use exactly the same models from \lemon as the input sources for our mutation techniques. The reason why we cannot use the our collected models for \tf 2 is because \tf 2 models cannot successfully execute on \tf 1.14. To show the importance of different sources, we also collect developer tests and documentation code for \tf v1.14. So the column `\tech (tf1.14 full)' in Table~\ref{tab:coverinput} is composed of the same models from \lemon, documentation code and tests. We can conclude that our code collected from diverse sources can cover more APIs and achieve higher coverage.}

\Comment{\cradle does not have mutation strategies, so we only compare with \lemon on mutation. We follow the same methodology as \lemon~\cite{lemon} when running model-level mutation. For \tech{}, we run 1000 mutants for each covered API. Note that the column `\tech models only' means that we only use the models from \lemon (without documentation code and developer tests) as the input to our mutation. From Table~\ref{tab:mutatecomparisonlemon}, we can conclude that considering various input sources (developer tests and official documentation code) helps greatly in achieving higher coverage (as shown by column `\tech full'). Also, given exactly the same input sources, API-level mutation (column `\tech{} models only') is able to achieve a higher coverage with much shorter time compared with model-level fuzzing (column `\lemon').}

\subsection{RQ5: Bugs Detected}
For bug detection, we target \pt 1.8 and \tf 2.4, which are both officially released stable versions, with the default \tech{} setting, i.e., generating 1000 mutants for each API.
Note that we do not target \tf 1.14\Comment{though \tech{} also detected multiple bugs there} because \Comment{it is a deprecated version and }developers do not actively support it anymore.
\Comment{Furthermore, we also include the bug detection information for various \tech variants studied in Section~\ref{subsec:variant}.}\Comment{There are around 80\% crash bugs which are actually due to invalid input. The reason is that our mutation algorithm does not guarantee that the input is valid. We have a script to heuristically remove crash bugs which throw meaningful exceptions for invalid input, e.g., `ValueError', `InvalidArgumentError', etc.}
Table~\ref{tab: sumdetected} shows the detailed statistics about the real-world bugs detected by \tech{} and its various variants studied in Section~\ref{subsec:variant}. We can observe that \tech is able to detect \NumAllBugs bugs in total (with \NumPreviouslyUnknownConfirm already confirmed as previously unknown bugs\Comment{\lingming{this 42 should have all been confirmed? we shall make it clear; also given 43 is not that much diff from 42, we can remove the 43 confirmed bugs (only keep `42 confirmed as previously unknown bugs') to avoid confusion}}) for the two studied DL libraries\Comment{\NumConfirmedBugs of the reported bugs have been confirmed by the developers}, and \numFixedBugs of them have been fixed by the developers to date. Furthermore, we can also observe that each mutation strategy can help detect certain bugs that other mutation strategies fail to detect, further demonstrating the importance of all \tech mutation strategies. Lastly, of all the \NumAllBugs bugs detected by \tech, only one of them can be detected by\Comment{state-of-the-art} \lemon and \cradle.
\begin{table}[!htp]\centering
\caption{Summary of detected bugs}
\label{tab: sumdetected}

\scalebox{0.78}{
\begin{tabular}{|l|c|c|c|c|c||c|}
\hline
&&\multicolumn{4}{c|}{\tech}&Confirmed\\
\cline{3-6}
& \tech & -TypeMu & -RandMu & -DBMu & -AllMu & (Fixed)\\
\hline
\hline
PyTorch & 28 & 13 & 24 & 26 & 5 & \NumConfirmedBugsPt ( \numFixedPt) \\
TensorFlow & 21 & 20 & 5 & 20 & 2 &  \NumConfirmedBugsTf (\numFixedTf)\\
\hline
\end{tabular}
}
\end{table}

\Comment{
\david{We had better not show Table~\ref{tab: catdetected}, and describe the categorization in text.}\lingming{agree. we should also see if we can find more bugs, it is okay to just report.}
\begin{table}[!htp]\centering
\caption{Categorization of detected bugs\lingming{only 1 bug for performance and wrong computation? too low!}}\label{tab: catdetected}
\begin{tabular}{|l|c|c|c|}
\hline
& \# Crash & \# Wrong Results & \# Performance Bugs \\
\hline
\hline
Pytorch & 12 & 1 & 1\\
Tensorflow & 14 & - & -\\
\hline
\end{tabular}
\end{table}
}

\Comment{
\subsection{Low coverage of \tf APIs}
\tech covers 701 of the \numHookAPIsTftwo \tf APIs (36.84\%), which is already a great improvement over previous methods. However, the coverage is still very low. Given that we have run all the documentation examples, official unit tests, and over 100 diverse models, the 1200 unused APIs must be very rarely used in practice, and we don't prioritize the testing of such unimportant APIs.
}
Note that all the detailed issue IDs for the bugs detected can be found on our GitHub page~\cite{freefuzzrepo}.
We next present the case studies:\Comment{ about some example bugs found by \tech:}

\parabf{Wrong-computation Bug.}
Figure~\ref{fig:difftestconv} shows an example bug automatically detected by \tech{} by comparing the computation results for \convtwod between two backends, one with CuDNN enabled (\CodeIn{output1}) and one disabled, using Aten backend instead (\CodeIn{output2}). It throws \CodeIn{AssertionError} when executing the last line. The sum of values of output tensors in Line~\ref{line:diffdebugging} shows that \CodeIn{output1 = -523.5300} while \CodeIn{output2 = -601.6165}, indicating a significant difference in computation results. Further comparing the computation results executed by CPU demonstrates that the result is wrong only on GPU with CuDNN disabled. This is a confirmed bug by developers and fixed in latest master.\Comment{, and the nightly build versions do not suffer from this anymore.}
\begin{figure}
	\begin{lstlisting}[basicstyle=\ttfamily\footnotesize,escapeinside={(*@}{@*)},columns=fixed,xleftmargin=3.5ex,numbers=left,language=Python]
m = torch.nn.Conv2d(64,128,1,2).cuda()
tensor = torch.rand(1,64,32,32).cuda()
torch.backends.cudnn.enabled = True
output1 = m(tensor) # with CuDNN enabled
torch.backends.cudnn.enabled = False
output2 = m(tensor) # with CuDNN disabled
print(output1.sum(), output2.sum()) # debugging(*@\label{line:diffdebugging}@*)
assert torch.allclose(output1, output2) # fail
    \end{lstlisting}
	\caption{Differential testing for \convtwod}
	\label{fig:difftestconv}
\end{figure}

\parabf{Performance bug.}
\tech{} detects one performance bug by metamorphic testing for \CodeIn{torch.nn.functional.conv\_transpose2d}. According to the metamorphic relations, the time cost for \textit{float16} computation should be less than that for \textit{float32} given the same parameters and tensor shapes. However, our\Comment{ testing} results on NVIDIA A100 GPU\Comment{(with CuDNN enabled)} for \pt show that \CodeIn{float16: cost = 0.377s, float32: cost = 0.101s} on some inputs. The bug detected by \tech{} has spurred a heated discussion among \pt developers. They confirmed this performance bug\Comment{ by running the bug-triggering script on different hardware themselves (2080ti, 1080ti, 1070 ti, etc). Developers appreciate our posting this interesting bug,} and are trying hard to figure out the reason\Comment{ for this}.

\parabf{Crash bug.}
\Comment{lingming, please choose one of the 2 crash bugs just below}
Figure~\ref{fig:crashconv3dpt} shows a crash bug detected by \tech{}\Comment{ with database value mutation strategy}. The program crashes on Line 3 when invoking\Comment{ API} \CodeIn{torch.nn.Conv3d}. The reason is that argument \CodeIn{padding\_mode} is set to value \CodeIn{`reflect'} and the program will not crash if \CodeIn{padding\_mode} is set to its default value `zeros'. The bug is triggered by the database mutation strategy. The argument name \CodeIn{padding\_mode} of type string appears in the \argdb, and there exists a value \CodeIn{`reflect'}, which is originally recorded for the argument \CodeIn{padding\_mode} of \CodeIn{torch.nn.Conv2d}. \Comment{Since the two APIs (\CodeIn{Conv2d} and \CodeIn{Conv3d}) are quite similar, }\tech{} applies the database mutation strategy to query the \argdb, and selects \CodeIn{`reflect'} from \CodeIn{Conv2d} to serve as the input for argument \CodeIn{padding\_mode} of \CodeIn{Conv3d}. We confirm this bug according to the documentation of \CodeIn{torch.nn.Conv3d}~\cite{conv3d_website_pt} where 4 string values (i.e., \CodeIn{`zeros'}, \CodeIn{`reflect'}, \CodeIn{`replicate'} or \CodeIn{`circular'}) should be valid for \CodeIn{padding\_mode}. Developers have acknowledged this bug and triaged it.
\begin{figure}
	\begin{lstlisting}[basicstyle=\ttfamily\footnotesize,escapeinside={(*@}{@*)},columns=fixed,xleftmargin=3.5ex,numbers=left,language=Python]
from torch.nn import Conv3d
x = torch.rand(2,3,3,3,3)
Conv3d(3,4,3,padding_mode='reflect')(x) # Crash
    \end{lstlisting}
	\caption{Crash bug in Conv3d}
	\label{fig:crashconv3dpt}
\end{figure}

\begin{figure}
	\begin{lstlisting}[basicstyle=\ttfamily\footnotesize,escapeinside={(*@}{@*)},columns=fixed,xleftmargin=3.5ex,numbers=left,language=Python]
m_gpu = torch.nn.MaxUnpool2d(2,stride=2).cuda()
m_cpu = torch.nn.MaxUnpool2d(2,stride=2)
tensor = torch.rand(1, 1, 2, 2)
indices = torch.randint(-32768,32768,(1, 1, 2, 2))
gpu_result = m_gpu(tensor.cuda(), indices.cuda())
cpu_result = cpu(tensor, indices) # only crash on CPU

    \end{lstlisting}
	\caption{Invalid test input for \CodeIn{torch.nn.MaxUnpool2d}}
	\label{fig:invalidinput}
\end{figure}

\Comment{Figure~\ref{fig:crashconv2dtranstf} presents a crash bug in \tf detected by \tech with database value mutation strategy. The program crashes on Line 3 when invoking the \CodeIn{Transpose2D} API. The root cause for this is that the argument \CodeIn{dilation\_rate} is set to a list of two different integers \CodeIn{(1, 2)} and the program will not crash if \CodeIn{dilation\_rate} is set to its default value \CodeIn{(1, 1)}. According to the documentation of \CodeIn{Conv2DTranspose}~\cite{conv2dtranspose_website_tf}, the \CodeIn{dilation\_rate} specifies the dilation rate for dilated convolution\Comment{and can be an integer or a list or tuple of two positive integers}. The bug is triggered due to the construction of \argdb and database value mutation. The value \CodeIn{(1,2)} is originally collected in the \CodeIn{Conv2D} \apivs from developer tests, and is selected from the \argdb of the value name \CodeIn{dilation\_rate} to replace the default value by the mutation algorithm, as these two APIs are quite similar. We confirm this bug by referring to the documentation of \CodeIn{Transpose2D}  \cite{conv2dtranspose_website_tf}, which claimed that \CodeIn{dilation\_rate} can be a list of two integers.

\begin{figure}[htb]
	\begin{lstlisting}[basicstyle=\ttfamily\footnotesize,escapeinside={(*@}{@*)},columns=fixed,xleftmargin=3.5ex,numbers=left,language=Python]
import tensorflow as tf
from keras.layers import Conv2DTranspose
x = tf.keras.Input([4,4,16])
Conv2DTranspose(1,32,dilation_rate=(1,2))(x) # Crash
    \end{lstlisting}
	\caption{Crash bug in Conv2DTranspose}
	\label{fig:crashconv2dtranstf}
\end{figure}}

\Comment{
\begin{figure}[htb]
	\begin{lstlisting}[basicstyle=\ttfamily\footnotesize,escapeinside={(*@}{@*)},columns=fixed,xleftmargin=3.5ex,numbers=left,language=Python]
import tensorflow as tf
x = tf.random.uniform((2, 8, 8, 8))
y = tf.keras.layers.ConvLSTM2D(0,3)(x) # Crash
    \end{lstlisting}
	\caption{Crash bug in ConvLSTM2D API from \tf}
	\label{fig:crashconvlstm2dtf}
\end{figure}

Figure~\ref{fig:crashconvlstm2dtf} shows a previously unknown crash bug in the \CodeIn{ConvLSTM2D} API from \tf by setting the argument \CodeIn{filters} to \CodeIn{0}. The program crashes without throwing any meaningful information when running the script shown in Figure~\ref{fig:crashconvlstm2dtf}. This bug has been confirmed and fixed by developers by raising \CodeIn{ValueError} exception because \CodeIn{0} is not a valid input as the number of filters.
}
\Comment{
\lingming{we should just keep one crash bug example due to space limit (comment out examples and dont remove them), better someone where db is helpful.}\tech detects several crash bugs. Figure~\ref{fig:testtorchsigmoid} shows the code to trigger the bug. It first initializes a tensor with \CodeIn{torch.complex64} as its data type. Then the execution result on CPU is successful, printing \CodeIn{tensor(1.1541+0.0255j)}, but fails when running on GPU. It throws RuntimeError: \CodeIn{"sigmoid\_cuda" not implemented for "ComplexFloat"}. The tensor with \textit{complex64} data type is recorded in the database\lingming{this is not argument database right? if i'm right, don't call it database to avoid confusion; also I don't know how db mutation helps for this case} of the parameter value space when running \pt official library tests (not specifically designed for \CodeIn{torch.sigmoid}, but in a testing function called \CodeIn{test\_all\_reduce\_sum\_complex}) for other APIs. This bug shows that it is beneficial to run differential testing because the bug can only be triggered on GPU. Moreover, the type mutation of \textit{\tensordtype} helps greatly, because although \CodeIn{torch.sigmoid} is well-tested for \textit{float32}, yet it crashes for \textit{complex64}. Also, the construction of \argdb
and the \textit{\dbtensorvalue} mutation strategy plays an important role in initializing tensor values from the \argdb from the database.
The developers appreciate our contribution in reporting this issue, and they have fixed this bug already by adding the complex data type support to \CodeIn{torch.sigmoid} on GPU.

\begin{figure}[htb]
	\begin{lstlisting}[basicstyle=\ttfamily\footnotesize,escapeinside={(*@}{@*)},columns=fixed,xleftmargin=3.5ex,numbers=left,language=Python]
import torch
t = torch.tensor(complex(3, 2))
print(torch.sigmoid(t)) # CPU: 1.1541+0.0255j
torch.sigmoid(t.cuda()) # GPU: RuntimeError
\end{lstlisting}
	\caption{Differential testing for torch.sigmoid}
	\label{fig:testtorchsigmoid}
\end{figure}

\Comment{
\tech{} detects another bug in \convthreed in a very interesting way. The script that can trigger the bug is shown in Figure~\ref{fig:testtorchconvthreed}. This script is almost the same as Figure~\ref{fig:conv-code} except that the \convthreed expects a 5-dimensional tensor. \tech{} is able to detect this bug by intentionally setting \CodeIn{padding\_mode="reflect"} when invoking \CodeIn{torch.nn.Conv3d} (the default parameter `zeros' cannot trigger the bug). It raises an exception \CodeIn{NotImplementedError}, but the official documentation for \convthreed says that ``\textit{padding\_mode (string, optional):  `zeros', `reflect', `replicate' or `circular'. Default: `zeros'} ", and the developers have also confirmed this. Interestingly enough, \tech{} detects this bug with the help of \textit{Database Primitive} value mutation strategy. Although no code from documentation, tests, or models sets `reflect' for \CodeIn{torch.nn.Conv3d} API, the specific string `reflect' is recorded in the database of argument value space during the execution of another API \CodeIn{torch.nn.functional.pad} (setting \CodeIn{mode = "reflect"}). The argument names of \CodeIn{mode} and \CodeIn{padding\_mode} is considered similar (based on text similarity). The bug is triggered by applying the \textit{Database Primitive} value mutation strategy (for string type).

\begin{figure}[htb]
	\begin{lstlisting}[basicstyle=\ttfamily\footnotesize,escapeinside={(*@}{@*)},columns=fixed,xleftmargin=3.5ex,numbers=left,language=Python]
import torch
m = torch.nn.Conv3d(3, 4, 3, padding_mode="reflect")
input = torch.rand(2, 3, 3, 3, 3)
output = m(input) # NotImplementedError
\end{lstlisting}
	\caption{Mutation testing for \convthreed}
	\label{fig:testtorchconvthreed}
\end{figure}
}

Our last example for crash bug comes from  \CodeIn{tf.math.acos} in \tf by applying \textit{Tensor Dtype} type mutation strategy. The code to trigger the bug is shown in Figure~\ref{fig:testtfacos}. It crashes with message ``NotFoundError: Could not find device for node...". The function output is normal with a \CodeIn{float32} input, but it crashes with \CodeIn{tf.half} (float16) tensors. Data type \CodeIn{half} should be supported according to the documentation of the API, and is indeed supported in an older version \tf 2.3.1, however, it is not supported in \tf 2.4.2. The bug is interesting in that it is a regression issue, which has been confirmed by developers.

\begin{figure}[htb]
	\begin{lstlisting}[basicstyle=\ttfamily\footnotesize,escapeinside={(*@}{@*)},columns=fixed,xleftmargin=3.5ex,numbers=left,language=Python]
import tensorflow as tf
a = tf.constant([1.0], dtype = tf.half)
tf.math.acos(a) # NotFoundError
\end{lstlisting}
	\caption{Mutation testing for acos from \tf}
	\label{fig:testtfacos}
\end{figure}

}

\Comment{\subsection{Future work}
\begin{itemize}
    \item Coverage-guided fuzzing
    \item A sequence of APIs
    \item Value-dependent APIs
    \item Generalization to numerical libraries
    \item Fuzzing augmented with automated reasoning from error message
    \item Model Parallelization correctness
\end{itemize}
}

\subsection{Threats to validity}
The threats to internal validity mainly lie in the correctness of the implementation of our own approach and the compared techniques. To reduce such threats, the\Comment{ first three} authors worked together to perform\Comment{extensive} testing and code review of \tech; also, we obtained the\Comment{original} implementation of\Comment{the} prior work from the official websites\Comment{ and authors}.

The threats to external validity mainly lie in the evaluation benchmarks used. To demonstrate that our \tech can be applied/generalized to different DL libraries, we have evaluated \tech on two most widely used DL libraries, \pt and \tf. Furthermore, although \tech is fuzzing against 1158 APIs (each with 1000 times) and the randomness can be largely mitigated by such a large number of APIs, it is still possible that the nondeterminism in \tech can affect the effectiveness of \tech in different runs~\cite{arcuri2011practical,klees2018evaluating}. Therefore, following existing fuzzing work~\cite{zhong2020squirrel,she2020mtfuzz,wen2020memlock}, we rerun the experimental comparison between \tech and \Comment{state-of-the-art}\lemon (Table~\ref{tab:mutatecomparisonlemon}) for 5 times. The results show that \tech achieves an average line coverage of 35997 (\linecovFreeComp in Table~\ref{tab:mutatecomparisonlemon}), while LEMON’s average is 29769 (\linecovlemon in Table~\ref{tab:mutatecomparisonlemon}). Both are quite stable with the coefficient of variation of only 0.82\%/0.06\%, demonstrating the effectiveness of \tech in different runs. 

\Comment{\david{
Due to the randomness of our fuzzing algorithm and the nondeterminism in \lemon, we run the comparison with \lemon on mutation (Table~\ref{tab:mutatecomparisonlemon}) for 5 times~\cite{zhong2020squirrel,she2020mtfuzz,wen2020memlock} to gain statistical confidence for better evaluation~\cite{arcuri2011practical,klees2018evaluating}.
Across 5 runs, Freefuzz’s average line coverage is 35997 (\linecovFreeComp in Table~\ref{tab:mutatecomparisonlemon}), while LEMON’s average is 29769 (\linecovlemon in Table~\ref{tab:mutatecomparisonlemon}). Both are quite stable with the coefficient of variation of only 0.82\%/0.06\% in line coverage.
}}

\Comment{The threats to external validity mainly lie in the open source models used in our experiments. To reduce this threat, we choose popular models from Github. During model selection, we abandon crashing models so that all our collected models do not originally contain crash bugs. Some models crash due to API updating, missing data, etc.}

\Comment{
\david{One thing that is worth mentioning is that developer tests may actually cover many more lines because \tech only instruments the public Python APIs. So the C++ tests and tests that target internal/private Python APIs do not count in line coverage in Section~\ref{subsec:inputsourcestudy}.}\lingming{not sure why you want to convey here, are we going to add this? why? fine with me if we remove this due to space limit}
}

The threats to construct validity mainly lie in the metrics used. To reduce such threats, we adopt the number of detected bugs used by prior work on DL library testing. Furthermore, we also include the widely used code coverage metric in traditional software testing.\Comment{, e.g., the number of bugs detected and code coverage. \lingming{can we say that we are the first one to use code coverage for testing DL libs?}\david{yes}}


\section{Discussion and Future Work}

\Comment{\parabf{Generalizability and Specificity.}
Different from \lemon~\cite{lemon} and \cradle~\cite{cradle} which only use DL models to test DL libraries, the basic idea of \tech{} can be generalized to more than just DL libraries, and we hope our work can inspire more research on the idea of fuzzing from mining.
We believe that leveraging the code snippets from library documentation and developer tests can be generalized to fuzzing APIs exposed in a dynamically typed language for many other libraries. With code instrumentation, one can obtain dynamic information on API arguments. With a set of carefully-design of mutation rules and proper oracles, automatic test generation is not difficult.
Meanwhile, we do have DL-specific components, including 1) mining DL models as inputs, 2) tensor-related types and mutation rules, 3) DL-specific oracles (e.g., for performance bugs).}

\parabf{Generalizability and Specificity.}
Different from \lemon~\cite{lemon} and \cradle~\cite{cradle} that specifically target testing DL libraries via DL models, the \tech{} work can potentially be generalized to more than just DL library fuzzing. Of course, in this work, we do have various DL-specific components, including 1) mining DL models as inputs, 2) tensor-related types and mutation rules, and 3) DL-specific oracles (i.e., differential testing for wrong-computation bugs and metamorphic testing for performance bugs). Meanwhile, the basic idea of leveraging code snippets from library documentation and developer tests can be generalized to fuzzing library APIs in various dynamically typed languages.\Comment{ With code instrumentation, one can obtain dynamic information on API arguments. With a set of carefully-design of mutation rules and proper oracles, automatic test generation is not difficult.}
We hope our work can inspire more research on the direction of mining for fuzzing.

\parabf{Validity of Test Inputs.}
Our input mining and type-aware~\cite{park2021generative}/DB-based mutations can all help generate more valid inputs. Meanwhile, \tech still does not always generate valid inputs due to some complicated input constraints. Interestingly, even the invalid inputs helped detect various bugs in \pt/\tf (e.g., unexpected crashes). Figure~\ref{fig:invalidinput} shows one such bug detected in \CodeIn{torch.nn.MaxUnpool2d}. The input parameter \CodeIn{indices} is a tensor whose values are randomly sampled integers (with respect to the \randomprimitive strategy), which is invalid. According to the documentation, the valid \CodeIn{indices} should be obtained from the returned value of \CodeIn{torch.nn.MaxPool2d}. The bug was detected because the program only crashes when running on CPU (i.e., Line 6 fails) but produces a wrong result silently without throwing any error message on GPU (i.e., Line 5 passes). Thus, the GPU implementation should add the missing check. The developers have confirmed this bug and even labelled with ``high priority\Comment{ for silent wrong result}''~\cite{bugreportMaxUnpool2d}.

\parabf{Future Work}.
\tech currently only focuses on testing the correctness of single APIs. While API-level testing has many advantages over model-level testing, it may still miss bugs which can only be triggered by invoking a sequence of APIs.\Comment{ Future work may explore how to detect bugs which can be triggered by a sequence of DL-library APIs.} Besides, when reproducing detected bugs, we find that some tests will fail on one machine but pass on other machines given exactly the same script and the same library version. This is probably due to the differences in underlying\Comment{ environment,} infrastructure and hardware. This type of tests are called implementation-dependent flaky tests, described in prior work on test flakiness~\cite{lam2020large,zhang2021domain,parry2021survey}. Future work may\Comment{ also} explore how to better detect and fix flaky tests~\cite{dutta2018testing,dutta2019storm,dutta2020detecting,dutta2021flex} in deep learning libraries.

\Comment{For example, an important  graph-level optimization in DL libraries is called operator fusion, which requires invoking multiple APIs as a sequence. In order to generate bug-triggering API sequences, future work can look into modelling the input-output relations of each API. Besides, the oracle part can still be improved. Some APIs do not have GPU support, and we hope that our work can enable more research on the oracle end. Last but not least, \tech{} does not test the correctness of back propagation for APIs. Back propagation (in other words, automatic differentiation) is the foundation for training a network and an indispensable component of a DL library. It will be of great interest to find bugs in the automatic differentiation engines for DL libraries.
}

\section{Related Work}

\Comment{\subsection{Fuzz testing}
\lingming{may be we don't need that}}

\parabf{DL Model Testing.}
There has been a growing body of research for improving the quality of DL models.
\Comment{
\parabf{Adversarial Robustness.}}
Various adversarial attack techniques~\cite{goodfellow2014explaining,moosavi2016deepfool,papernot2016limitations,zhang2021advdoor} have been proposed to generate the so-called ``adversarial examples'' by adding perturbations imperceptible to humans to intentionally fool the classification results given by DL models. To mitigate such attacks, researchers have also proposed various adversarial defense techniques, including adversarial training~\cite{madry2017towards,goodfellow2014explaining,szegedy2013intriguing}, detection~\cite{ma2018characterizing,gu2019detecting,zhao2021attack}, and others~\cite{wang2019adversarial}.
Another recent line of research has explored the possibility of improving the robustness of neural network from a joint perspective of traditional software testing and the new scenario of deep learning. DeepXplore~\cite{pei2017deepxplore} proposes a metric called neuron coverage for whitebox testing of DL models and leveraged gradient-based techniques to search for more effective tests. While various other metrics~\cite{kim2019guiding,ma2018deepgauge} have also been proposed recently, the correlation between such metrics and the robustness of models is still unclear~\cite{yan2020correlations,dong2020empirical,harel2020neuron}. Meanwhile, there are also a series of work targeting specific applications, such as autonomous driving, including DeepTest~\cite{tian2018deeptest}, DeepRoad~\cite{zhang2018deeproad}, and DeepBillboard~\cite{zhou2020deepbillboard}. Various techniques have also been proposed to detect numerical bugs introduced when building a DNN model at the architecture level with static analysis~\cite{zhang2020detecting}, and via gradient propagation~\cite{yan2021exposing}. Lastly, researchers have also explored concolic testing~\cite{sun2018concolic} to achieve higher coverage for DNN models, mutation testing~\cite{ma2018deepmutation,humbatova2021deepcrime} to simulate real faults in DL models, and test input generation for DNNs~\cite{dunn2021exposing} by exploiting features learned from data with generative machine learning. Different from all such prior work, our work targets the underlying DL libraries, which are the basis for training and deploying various DL models. 

\Comment{\david{I think DeepMutation is irrelevant.}

\lingming{deepxplore, deeptest, deeproad, deepbillboard, deepgauge, deepmutation, deep...}}

\parabf{DL Library Testing.}
\cradle~\cite{cradle} is the trailblazing work for testing DL libraries. The main contribution of \cradle is resolving the test oracle challenge with differential testing on \Comment{different backends of }\keras. \lemon~\cite{lemon} further advanced \cradle{} by proposing mutation rules to generate more models, as claimed by \lemon to invoke more code in DL libraries. \lemon's mutation strategies include intact-layer and inner-layer mutation rules, which must conform to strict constraints, e.g., for intact-layer mutation, the layer to be inserted or deleted should preserve the shapes of input tensors.\Comment{ However, some of the backends (e.g., Theano) are no longer maintained by developers~\cite{theano_abandoned}.} Actually, according to our experimental results, the mutation rules applied by \lemon can hardly help cover more DL library code. A more recent work on testing DL library is \predoo~\cite{zhang2021predoo}, which only mutates the input tensor values with all other API parameters \emph{manually} set up for precision testing. As a result, it was only applied to \predooOperatorNum APIs/operators from \tf and we exclude it in our comparison.\Comment{ \lingming{what manual efforts needed for this tech, they can only test 7 apis? also, any more recent techniques to be discussed?}\david{\predoo only mutates the input tensor values with all other API parameters manually set up}.} \Comment{FPDiff~\cite{vanover2020discovering} aims to discover meaningful discrepancies
across numerical libraries with differential testing.} To our knowledge, we propose the first general-purpose and fully automated API-level fuzzing approach for popular DL libraries. Furthermore, we adopt traditional code coverage for DL library testing, and reveal various interesting findings (e.g., state-of-the-art \lemon can hardly improve the DL library code coverage).

\Comment{\david{
Besides, due to the randomness in the algorithms and implementations of deep learning libraries~\cite{pham2020problems}, there is a growing body of research on detecting and fixing flaky tests for machine learning applications and probabilistic programming systems~\cite{dutta2018testing,dutta2019storm,dutta2020detecting,dutta2021flex}.
}\lingming{not sure how to elegantly include this, this is definitely not the right place; maybe we can just add their citations to the future work part and remove the description}
\david{I Agree. I have moved Sara's citations to Future Work.}}
\Comment{
None of the prior work use code coverage as the metric in their experiments.

We are the first to apply code coverage in our experiments, and we are the first to test both \pt and \tf.
\david{It raises concerns including how our coverage is compared with coverage of developer tests! We want to muddy the water.}

\lingming{junjie's new fse about numerical bugs}
\lingming{check the issta'21 program for some more dl lib testing}


\lingming{https://conf.researchr.org/program/issta-2021/program-issta-2021/. this will provide more related work for both dl lib and model testing}}

\section{Conclusion}

We have proposed \tech{}, the first approach to fuzzing DL libraries via mining from open source. More specifically, \tech{} considers three different sources: 1) library documentation, 2) developer tests, and 3) DL models in the wild. Then, \tech{} automatically runs all the collected code/models with instrumentation to trace the dynamic information for each covered API\Comment{, including the types and values of each parameter during invocation, and shapes of input and output tensors}. Lastly, \tech{} will leverage the traced dynamic information to perform fuzz testing for each covered API. 
The extensive study of \tech{} on \pt and \tf\Comment{,two of the most popular DL libraries, } shows that \tech{} is able to automatically trace valid dynamic information for fuzzing \numTracedAPIs\Comment{ of \numHookAPIs} popular APIs, \APImorecovered more than state-of-the-art \lemon with \overheadlowerfull lower overhead.\Comment{Furthermore, leveraging the traced information,} \tech{} has detected \NumAllBugs bugs for \pt and \tf (with \NumPreviouslyUnknownConfirm already confirmed by developers as previously unknown bugs).

\section*{Acknowledgments}
We thank Darko Marinov, Chenyang Yang, and Matthew Sotoudeh for their valuable discussions and suggestions\Comment{ on the paper}. We also appreciate the insightful comments from the anonymous reviewers. This work was partially supported by National Science Foundation
under Grant Nos. CCF-2131943 and CCF-2141474, as well as Ant Group.
\balance
\bibliographystyle{abbrv}
\bibliography{references}










\end{document}